 \documentclass[10pt,a4paper,english,journal, twocolumn]{IEEEtran}

 \usepackage[T1]{fontenc}
 \usepackage[latin9]{inputenc}
 \usepackage{float}
 \usepackage{amsmath}
 \usepackage{amssymb}
 \usepackage{times}
 \usepackage{amsfonts}
 \usepackage{graphicx}
 \usepackage{cite}
 \usepackage{threeparttable}
 \usepackage[color]{changebar}
 \cbcolor{red}
 \usepackage{picinpar}
 \usepackage{rotating}
 \usepackage{url}

  \usepackage{psfrag}
  \usepackage{placeins}

\usepackage{setspace}
 \usepackage{multicol}
 \usepackage{color}
 \usepackage[footnotesize]{caption}
 \usepackage{algorithm}
 \usepackage{algpseudocode}
 \makeatletter

 \floatstyle{ruled}
 \newfloat{algorithm}{tbp}{loa}
 \providecommand{\algorithmname}{Algorithm}
 \floatname{algorithm}{\protect\algorithmname}

 \input epsf

 \usepackage{babel}\addto\captionsbreton{\renewcommand{\figurename}{Fig.}}
 \addto\captionsbulgarian{\renewcommand{\figurename}{Fig.}}
 \addto\captionsenglish{\renewcommand{\figurename}{Fig.}}

 \@ifundefined{showcaptionsetup}{}{%
  \PassOptionsToPackage{caption=false}{subfig}}
 \usepackage{subfig}
 \makeatother

\begin{document}
\bibliographystyle{../../../../../shaoshi_bib/IEEEtran}
\title{Iterative Distributed Minimum Total-MSE Approach
for Secure Communications in MIMO Interference Channels }
\author{Zhengmin Kong, Shaoshi Yang, \IEEEmembership{Member,~IEEE}, Feilong
Wu, Shixin Peng, Liang Zhong, \\ and Lajos~Hanzo, \IEEEmembership{Fellow,~IEEE}\\
\thanks{Copyright (c) 2013 IEEE. Personal use of this material is permitted. However, permission to use this material for any other purposes must be obtained from the IEEE by sending a request to pubs-permissions@ieee.org.

The financial support of the National Natural Science Foundation of China (NSFC) (Grant 61201168) and of the European Research Council's Advanced Fellow Grant is gratefully acknowledged.

Z. Kong is with the Automation Department, Wuhan University, Wuhan 430072, China, and also with the School of Electronics and Computer Science, University of Southampton, Southampton SO17 1BJ, UK (e-mail: zmkong@whu.edu.cn).

S. Yang and L. Hanzo are with the School of Electronics and Computer Science, University of Southampton,
Southampton SO17 1BJ, UK (e-mail: \{sy7g09, lh\}@ecs.soton.ac.uk).

F. Wu is with the School of Electronics and Information, Xi'an Jiaotong University, Xi'an 710049, China, and also with the Xi'an Branch, China Academy of Space Technology (e-mail: wufeilong@stu.xjtu.edu.cn).

S. Peng is with State Grid Information and Communication Company of Hunan Electric Power Corp., Changsha 410007, China (e-mail: psx6050@163.com).

L. Zhong is with the Automation Department, Wuhan University, Wuhan 430072, China (e-mail: zhongliang@whu.edu.cn).
}}

\markboth{Accepted to appear on IEEE Transactions on Information Forensics \& Security, Oct. 2015}%
{Shell \MakeLowercase{\textit{et al.}}: Bare Demo of IEEEtran.cls
for Journals}

\maketitle
\vspace{-2cm}

\begin{abstract}
In this paper, we consider the problem of joint transmit precoding (TPC) matrix and receive filter matrix design subject to both secrecy and per-transmitter power constraints in the MIMO interference channel, where $K$ legitimate transmitter-receiver pairs communicate in the presence of an external eavesdropper. Explicitly, we jointly design the TPC and receive filter matrices based on the minimum total mean-squared error (MT-MSE) criterion under a given and feasible information-theoretic \textit{degrees of freedom}. More specifically, we formulate this problem by minimizing the total MSEs of the signals communicated between the legitimate transmitter-receiver pairs, whilst ensuring that the MSE of the signals decoded by the eavesdropper remains higher than a certain threshold. We demonstrate that the joint design of the TPC and receive filter matrices subject to both secrecy and transmit power constraints can be accomplished by an efficient iterative distributed algorithm. The convergence of the proposed iterative
algorithm is characterized as well. Furthermore, the performance of the proposed algorithm, including both its secrecy rate and MSE, is characterized with the aid of numerical results. We demonstrate that the proposed algorithm outperforms the traditional interference alignment (IA) algorithm in terms of both the achievable secrecy rate and the MSE. As a benefit, secure communications can be guaranteed by the proposed algorithm for the MIMO interference channel even in the presence of a ``sophisticated/strong'' eavesdropper, whose number of antennas is much higher than that of each legitimate transmitter and receiver.
\end{abstract}


\vspace{-0.2cm}

\section{Introduction}

\label{Section:Introduction}

Multiple-input-multiple-output (MIMO) techniques have attracted significant research interests over the past two decades due to their huge potential of boosting the spectral efficiency of wireless communication systems.
Recently, the MIMO interference channel, where $K$ transmitter-receiver pairs having multiple antennas at each entity communicate in
parallel, has become the centre of attention for the sake of maximizing the sum rate of an interference-contaminated peer-to-peer (P2P) multiuser network \cite{Jafar_DoF_GLOBECOM2007,Rahman_Mitigation_MIMO_Co-Channel_Interference_VTC2007,Gou_TIT2010,Mehana_MMSE_MIMO_receivers_GLOBECOM2013,Zeng_Rank_Deficient_DoF_TWC2014}. To elaborate a little further, in the MIMO interference channel, each of the $K$ multiple-antenna transmitters only wants to communicate with its dedicated multiple-antenna receiver, and each of the $K$ receivers only cares about the information arriving from the corresponding transmitter \cite{Yang_MIMO_PIEE2014,Carleial_IC_TIT1978}. As a result, each transmitter-receiver link is interfered by $K-1$  links, and in total there are $K$ principal links and $K(K-1)$ interference links constituting a MIMO interference channel.

Despite numerous significant advances in wireless communications, due to the distributed nature of both the transmitters and receivers \cite{Yang_MIMO_PIEE2014,Carleial_IC_TIT1978}, the management of the interference in the context of this interference channel still has numerous open facets. Exceptions exist only for a few special cases: i) when the interference is ``very weak'' \cite{Etkin_Gaussian_IC_TIT2008,Motahari_Capacity_IC_TIT2009,Shang_Bound_IC_TIT2009,Annapureddy_Capacity_low_interference_TIT2009}, treating the interference as noise is optimal, ii) and when the interference is ``very strong'' \cite{Carleial_Interference_not_Reduce_Capacity_TIT1975,Sato_Capacity_Strong_Interference_TIT1981,Han_Rate_Region_IC_TIT1981}, eliminating the interference with the aid of multiuser/MIMO detection\cite{Yang_MIMO_PIEE2014} is optimal\footnote{Note that the scenario where multiuser detection is applicable reduces to an interfering multiple-access channel, which is simpler than the interference channel.}.
This landscape has been changed, however, by a recent breakthrough, namely by the invention of interference alignment (IA) \cite{Maddah_Multi_Access&Broadcast_IC_ISIT2006,Maddah_IA_TIT2008,Cadambe_TIT2008}. The IA technique is proved optimal for the specific scenario where the interference power is comparable to that of the desired signal. With the aid of the IA technique, theoretically the sum-capacity of interference networks using limited time-, frequency- and spatial-domain radio resources, can be increased linearly with the number of users, which is in stark contrast to the conventional wisdom relying on the ``cake-cutting'' style of multiuser resource allocation\cite{Cadambe_TIT2008}. The basic idea of IA is that of maximizing the interference-free space for the desired signal, which is achieved by using carefully designed transmit beamforming/precoding vectors for each transmitter so that all the interference can be concentrated roughly into one half of the signal space at each receiver, leaving the other half available for the desired signal and free of interference \cite{jafar_2011:IA_tutorial,Heath_2013:IA_magazine,Cadambe_TIT2008,Gao_Limited_Feedback_IA_TIT2014,Cheung_Energy_Efficiency_IA_TSP2015}.

Although IA is an effective technique of managing the MIMO interference, 
when there is an external eavesdropper,
the resultant physical layer security problem remains a significant challenge \cite{Swindlehurst_SINR_MIMO_wiretap_channel_ICASSP2011,Cumanan_Secrecy_rate_TVT2014,Xie_Secure_DoF_one-hop_Networks_TIT2014, Pinto_2012:secure_comm_part_I,Pinto_2012:secure_comm_part_II,Zhang_2013:secrecy_MIMO_ad_hoc_2013}. More specifically, in a wiretap channel, when an external eavesdropper
is capable of obtaining the \textit{precoded} channel state information (CSI)
between itself and the legitimate transmitters\footnote{The precoded CSI, namely the product of the precoding matrix and the channel matrix, may be obtained by exploiting the pilots sent from the legitimate transmitters.}, secure communications may no longer be guaranteed.
More particularly, if the eavesdropper is ``sophisticated'', namely when the number of the eavesdropper
antennas is higher than that of each legitimate transmitter and receiver,
the secrecy rate of the user compromised  by the eavesdropper may reduce to zero, which implies that the communication of this user is insecure. Therefore, it is necessary to develop an IA scheme that is capable of guaranteeing secure wireless communications by explicitly
taking into account the secrecy constraints imposed by the eavesdropper.

The existing contributions on the joint study of IA and secure communications mainly focused on
the analysis and design of secure interference networks based on the information-theoretic concept of achieving a certain maximum degrees of freedom (DoF)
\cite{Koyluoglu_Secrecy_IA_TIT2011,Bassily_Ergodic_Secret_IA_TIT2012,Khisti_DoF_MISO_Wiretap_Channel_TIT2011,
Gou_Secure_DoF_Allerton2008}.
In contrast to these DoF-based studies, 
the authors of \cite{Fakoorian_Confidential_Interference_Channel_TIFS2011}
studied the achievable rate regions of the MIMO interference channel
where confidential messages are sent to two receivers. However, the analysis was limited to the two-user MIMO interference channel and no  eavesdropper was considered. Additionally, the authors of \cite{Sasaki_Secure_IA_ISAP2012}
presented a quantitative evaluation of the secrecy rate that is achievable in
secure IA communications involving an eavesdropper. It was shown in \cite{Sasaki_Secure_IA_ISAP2012} that the traditional distributed IA scheme is capable of obtaining a useful \textit{positive} secrecy rate in the MIMO interference channel under a very special condition, where the number of antennas
of each legitimate transmitter and receiver is higher than or equal
to that of the eavesdropper. However, this approach fails to ensure secure communications in the challenging IA scenario, where
the number of antennas of each legitimate transmitter and receiver
is less than that of the eavesdropper.

To the best of our knowledge, there is no open literature addressing the transceiver analysis and design relying on the joint optimization of the transmit precoding (TPC) matrix and receive filter matrix conceived for secure communications
in the presence of an eavesdropper over the \textit{MIMO interference channel}.
Motivated by this challenge, in this paper we aim for providing an algorithm for the minimization of the total mean-square error (MSE) in order to jointly design the TPC and receive filter matrices.

The MSE-based transceiver design has been widely investigated both in traditional point-to-point and in multiuser MIMO systems   \cite{Chang_MSE_Joint_transceiver_TWC2002,Schubert_MSE_SINR_constraints_TVT2004,Shi_MMSE_optimization_ICC2008,
Shi_MMSE_multiuser_MIMO_TSP2008,Tenenbaum_MSE_multiuser_MIMO_TWC2009,
Mehana_MMSE_MIMO_receivers_GLOBECOM2013}.
Additionally, the authors of  \cite{Peters_TVT2011,Shen_MSE_transceiver_TWC2010,Torabi_MSE_orthogonal_beamformer_IA_PIMRC2011,Shi_iteratively_weighted_MMSE_TSP2011,
Razaviyayn_Linear_transceiver_complexity_TIT2012,Zhang_Robust_per-stream_MSE_GLOBECOM2013,Kim_MSE_hybrid_RF_MIMO_interference_TVT2014}
have conceived dedicated algorithms based on MSE optimization both for the MIMO interference
channel and for the MIMO Gaussian wiretap channel\cite{Reboredo_sumMSE_transmitter_power_constraints_TSP_2013}. However, the secrecy constraints with respect to the eavesdropper
were not considered in \cite{Peters_TVT2011,Shen_MSE_transceiver_TWC2010,Torabi_MSE_orthogonal_beamformer_IA_PIMRC2011,Shi_iteratively_weighted_MMSE_TSP2011,Razaviyayn_Linear_transceiver_complexity_TIT2012,Zhang_Robust_per-stream_MSE_GLOBECOM2013,Kim_MSE_hybrid_RF_MIMO_interference_TVT2014}.
Hence, in this paper, we propose an iterative distributed minimum total-MSE (MT-MSE) algorithm that jointly designs the TPC and receive filter matrices for secure communications over the
MIMO interference channel in the presence of an external eavesdropper. This MT-MSE algorithm is capable of striking a
tradeoff between the grade of security and the achievable interference mitigation for the MIMO interference
channel in multiple aspects. In particular, it is capable of handling the minimization of the total MSE of the recovered signals, while maintaining secure communications over the MIMO interference channel. Furthermore,
relying on the proposed MT-MSE algorithm we demonstrate that a useful positive secrecy rate
can be guaranteed even for the challenging scenario, where there is a sophisticated eavesdropper that has a higher number of
antennas than each legitimate transmitter and receiver. Additionally, both our analytical and numerical results show that the proposed MT-MSE algorithm converges after a few iterations to a constant.

The remainder of the paper is organized as follows. In Section II we describe
the system model. In Section III, we propose our MT-MSE algorithm
for secure communications in the IA-aided MIMO interference channel. The convergence of our algorithm is characterized in Section IV, while numerical results are provided in Section V for characterizing the performance of the proposed algorithm. Finally, our conclusions are provided in Section VI.

\textit{Notation:} Throughout this paper, we use $\left(\cdot\right)^{*}$ to represent conjugate, $\left(\cdot\right)^{H}$  Hermitian transpose (i.e. conjugate transpose), $\mathrm{Tr}\left(\cdot\right)$ the trace of a matrix, $\mathbb{E}\left\{ \cdot\right\} $ the expectation, $\left\Vert \,\cdot\,\right\Vert _{F}$ the Frobenius norm of a matrix, $\left\Vert \,\cdot\,\right\Vert $ the 2-norm of a vector, $\left|\,\cdot\,\right|$ the determinant, $\mathbf{I}$ the identity matrix, and $\mathbf{0}$ a matrix or vector whose all entries are zeros. Furthermore, we use uppercase bold-face letters for denoting matrices, while keeping the lowercase bold-face for vectors and lowercase normal-face for scalars. $\nabla f\left(\cdot\right)$ is invoked to denote the gradient of the function $f\left(\cdot\right)$. $\mathcal{CN}\left(0,1\right)$ represents the complex circularly symmetric Gaussian distribution with zero mean and unit variance. $\mathbb{C}$  denotes the complex field.

\section{System Model}

\label{Section:System_Model}
\begin{figure}
\begin{centering}
\includegraphics[width=3.5in]{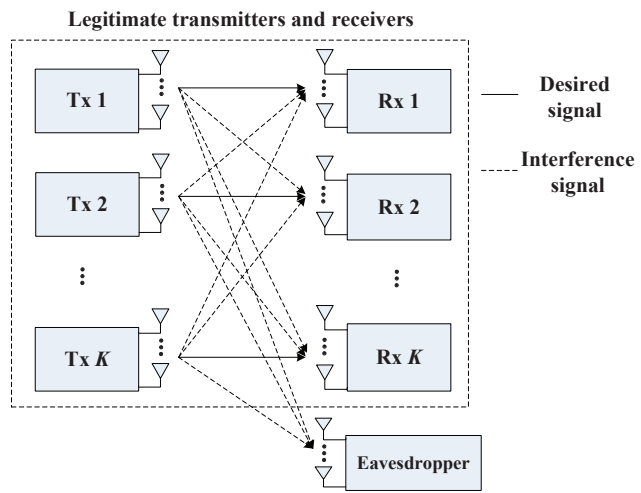}
\par\end{centering}
\centering{}\protect\caption{$K$-user MIMO interference channels with an eavesdropper.}
\label{fig:system_model}
\end{figure}

In this paper, we consider the scenario of $K$-user MIMO interference
channels in the presence of an eavesdropper, as seen in Fig. \ref{fig:system_model}, where the transmitter $k$, the receiver
$k$ and the eavesdropper are equipped with $N_{k}$, $M_{k}$ and
$M_{e}$ antennas, respectively, $k = 1, \cdots, K$. We have the following system model
from the perspective of Receiver $k$ and the eavesdropper:
\begin{eqnarray}
\mathbf{y}_{k}=\sum_{l=1}^{K}\mathbf{H}_{kl}\mathbf{x}_{l}+\mathbf{z}_{k},\label{eq:channel_output_at_receiver_k}
\end{eqnarray}
\begin{eqnarray}
\mathbf{y}_{e}=\sum_{l=1}^{K}\mathbf{H}_{el}\mathbf{x}_{l}+\mathbf{z}_{e},\label{eq:channel_output_at_Eve}
\end{eqnarray}
where $\mathbf{y}_{k} \in \mathbb{C}^{M_k \times 1}$ is the
received signal vector at Receiver $k$; $\mathbf{y}_{e} \in \mathbb{C}^{M_e \times 1}$ is the
received signal vector at the eavesdropper; $\mathbf{x}_{l} \in \mathbb{C}^{N_l \times 1}$ is the
transmitted signal vector at Transmitter $l$; $\mathbf{H}_{kl} \in \mathbb{C}^{M_k \times N_l}$ is the
matrix of channel coefficients between Transmitter $l$ and Receiver
$k$; $\mathbf{H}_{el} \in \mathbb{C}^{M_e \times N_l}$ is the matrix of channel coefficients between Transmitter $l$ and the eavesdropper, $\mathbf{z}_{k} \in \mathbb{C}^{M_k \times 1}$ is the complex additive white Gaussian noise (AWGN) vector at Receiver $k$ with zero mean and covariance matrix $\sigma_{k}^{2}\mathbf{I}$,
i.e. we have $\mathbf{z}_{k}\sim\mathcal{CN}\left(\mathbf{0},\sigma_{k}^{2}\mathbf{I}\right)$;
$\mathbf{z}_{e} \in \mathbb{C}^{M_e \times 1}$ is the complex AWGN vector at the eavesdropper, i.e. $\mathbf{z}_{e}\sim\mathcal{CN}\left(\mathbf{0},\sigma_{e}^{2}\mathbf{I}\right)$.

To realize interference-free communications in these $K$ legitimate
transmitter-receiver pairs, IA is employed. On the transmitter
side, Transmitter $k$ uses the TPC matrix $\mathbf{V}_{k} \in \mathbb{C}^{N_k \times d_k}$ to map the $d_{k}$ data
symbols contained in $\mathbf{s}_{k}$ to its $N_{k}$ transmit
antennas according to
\begin{eqnarray}
\mathbf{x}_{k}=\mathbf{V}_{k}\mathbf{s}_{k},\label{eq:transmit_with_TPC}
\end{eqnarray}
where the transmitted data symbols are independent and identically
distributed (i.i.d.) so that we have $\mathbb{E}\left\{ \mathbf{s}_{k}\mathbf{s}_{k}^{H}\right\} =\mathbf{I}$
and $\mathbb{E}\left\{ \mathbf{s}_{k}\mathbf{s}_{l}^{H}\right\} =\mathbf{0}$
($\forall k\neq l$); the TPC matrix $\mathbf{V}_{k}$ having linearly independent columns is normalized to ensure that $\parallel\mathbf{V}_{k}\parallel_{F}^{2}\leq\mathbf{\mathrm{\mathit{p_{k}}}}$, where
$\mathbf{\mathrm{\mathit{p_{k}}}}$ is the transmit power at Transmitter
$k$. Rewriting (\ref{eq:channel_output_at_receiver_k}) and (\ref{eq:channel_output_at_Eve}),
Receiver $k$ and the eavesdropper respectively observe the signals
\begin{eqnarray}
\mathbf{y}_{k} 
 & = & \mathbf{H}_{kk}\mathbf{V}_{k}\mathbf{s}_{k}+\sum_{l\neq k}^{K}\mathbf{H}_{kl}\mathbf{V}_{l}\mathbf{s}_{l}+\mathbf{z}_{k},\label{eq:channel_output_at_receiver_k_with_TPC} \\
\mathbf{y}_{e}& = & \sum_{l=1}^{K}\mathbf{H}_{el}\mathbf{V}_{l}\mathbf{s}_{l}+\mathbf{z}_{e}.\label{eq:channel_output_at_Eve_with_TPC}
\end{eqnarray}

On the receiver side, Receiver $k$ uses the receive filter matrix $\mathbf{U}_{k} \in \mathbb{C}^{d_k \times M_k}$ to obtain the estimate of the data symbols
$\mathbf{s}_{k}$ according to:
\begin{eqnarray}
\mathbf{\hat{s}}_{k}=\mathbf{U}_{k}^{H}\mathbf{H}_{kk}\mathbf{V}_{k}\mathbf{s}_{k}+\mathbf{U}_{k}^{H}\sum_{l\neq k}^{K}\mathbf{H}_{kl}\mathbf{V}_{l}\mathbf{s}_{l}+\mathbf{U}_{k}^{H}\mathbf{z}_{k}.\label{eq:estimated_symbol_at_receiver_k}
\end{eqnarray}
Naturally, if no eavesdropper is present, these $K$ legitimate
transmitter-receiver pairs are capable of achieving their maximum communication
rates when an IA scheme is employed; meanwhile the communications
are reliable and secure. However, the reliable secure communications are no longer guaranteed in the presence of an eavesdropper, who is capable of compromising any of the transmitters' transmissions. Assuming that the eavesdropper can use the receive filter matrix $\mathbf{U}_{e,k}$ ($k\in\left\{ 1,2,\ldots,K\right\} $),
which is calculated by the eavesdropper relying on minimizing its own detection MSE\footnote{${\bf U}_k$ and ${\bf U}_{e,k}$ are typically different from each other, since Receiver $k$ and the eavesdropper calculated their receive filter matrices relying on minimizing their local detection MSE.}, the data symbols $\mathbf{s}_{k}$
intended by Transmitter $k$ to Receiver $k$ can be detected by the eavesdropper according to:
\begin{eqnarray}
\mathbf{\hat{s}}_{e,k}=\mathbf{U}_{e,k}^{H}\mathbf{H}_{kk}\mathbf{V}_{k}\mathbf{s}_{k}+\mathbf{U}_{e,k}^{H}\sum_{l\neq k}^{K}\mathbf{H}_{kl}\mathbf{V}_{l}\mathbf{s}_{l}+\mathbf{U}_{e,k}^{H}\mathbf{z}_{e}.\label{eq:estimated_symbol_at_Eve}
\end{eqnarray}
That is, when the eavesdropper is present, the TPC matrix $\mathbf{V}_{k}$ and the receive filter matrix $\mathbf{U}_{k}$ employed by Transmitter $k$ no longer guarantee maintaining
the secrecy rate at which transmitters can reliably send their secret messages to the receivers. Hence, the set of  the TPC and receive filter matrices $\left\{ \mathbf{V}_{k},\mathbf{U}_{k}\right\}$, $k\in\left\{ 1,2,\ldots,K\right\} $
must be recalculated. In order to realize secrecy communications, as stated in Section \ref{Section:Introduction}, we minimize the total MSE of the signals received at all legitimate receivers subject to the constraints that the MSE of the signal recovered at the eavesdropper must be larger than a given threshold and the maximum transmit power of each transmitter is not exceeded.

\section{Secure MT-MSE Algorithm For The MIMO Interference Channel}
\label{Section:Minimum Total MSE algorithm for secure IA}

In this section, we focus our attention on a common metric in order to avoid enhancing the noise in the linear detectors of MIMO systems, namely the MSE metric \cite{Yang_MIMO_PIEE2014, Peters_TVT2011}. According to our MT-MSE algorithm,
we jointly design the TPC and receive filter matrices to solve the
total MSE minimization problem in the presence of an external eavesdropper.
First, we describe the total MSE objective with secrecy and transmit power
constraints for the MIMO interference channel. Our second goal
is to find the first-order optimality conditions with respect to each
TPC and receive filter matrix-pair. Finally, we present our iterative
distributed approach (i.e. the MT-MSE algorithm) of obtaining the TPC
and receive filter matrices.

We concentrate on MSE minimization owing to its attractive performance-complexity tradeoff\cite{Yang_MIMO_PIEE2014}. The TPC and receive filter matrices $\left\{ \mathbf{V}_{k},\mathbf{U}_{k}\right\}$, $k\in\left\{ 1,2,\ldots,K\right\} $
will be designed so that the total MSE of all receivers is minimized.
Under the assumption of independence between  $\mathbf{s}_{k}$
and $\mathbf{z}_{k}$, the error covariance matrix
$\mathrm{\mathbf{E}_{Rx_{\mathit{k}},Tx_{\mathit{k}}}}$ at Receiver
$k$ can be written as \eqref{eq:MSE_cov_matrix_receiver_k}.
\begin{figure*}
\begin{align}
\mathrm{\mathbf{E}_{Rx_{\mathit{k}},Tx_{\mathit{k}}}}\!\triangleq \! \mathbb{E}\!\left\{ \left(\mathbf{\hat{s}}_{k}\!-\!\mathbf{s}_{k}\right)\left(\mathbf{\hat{s}}_{k}\!-\!\mathbf{s}_{k}\right)^{H}\right\}
\!=\! \mathbb{E}\!\left\{ \mathbf{U}_{k}^{H}\!\left(\sum_{l=1}^{K}\mathbf{H}_{kl}\mathbf{V}_{l}\mathbf{V}_{l}^{H}\mathbf{H}_{kl}^{H}\right)\mathbf{U}_{k}\!-\!\mathbf{U}_{k}^{H}\mathbf{H}_{kk}\mathbf{V}_{k} \!-\!\mathbf{V}_{k}^{H}\mathbf{H}_{kk}^{H}\mathbf{U}_{k}\!+\!\sigma_{k}^{2}\mathbf{U}_{k}^{H}\mathbf{U}_{k}\!+\!\mathbf{I}\right\}. \label{eq:MSE_cov_matrix_receiver_k}
\end{align}
\end{figure*}

Similarly, the error covariance matrix $\mathrm{\mathbf{E}_{Eve,Tx_{\mathit{k}}}}$
at the eavesdropper who aims for wiretapping the data symbols $\mathbf{s}_{k}$ sent from  Transmitter $k$ to Receiver $k$ is defined as \eqref{eq:MSE_cov_matrix_Eve}.
\begin{figure*}
\begin{align}
\mathrm{\mathbf{E}_{Eve,Tx_{\mathit{k}}}}\!\!\triangleq\! \mathbb{E}\!\left\{ \!\left(\mathbf{\hat{s}}_{e,k}\!\!-\!\mathbf{s}_{k}\!\right)\!\left(\mathbf{\hat{s}}_{e,k}\!\!-\!\mathbf{s}_{k}\!\right)^{H}\!\right\}
\!=\! \mathbb{E}\!\!\left\{\! \mathbf{U}_{e,k}^{H}\!\!\left(\sum_{i=1}^{K}\mathbf{H}_{ei}\mathbf{V}_{i}\mathbf{V}_{i}^{H}\mathbf{H}_{ei}^{H}\!\!\right)\!\!\mathbf{U}_{e,k}\!\!+\!\mathbf{I} \!-\!\mathbf{U}_{e,k}^{H}\mathbf{H}_{ek}\mathbf{V}_{k}\!\!-\!\mathbf{V}_{k}^{H}\mathbf{H}_{ek}^{H}\mathbf{U}_{e,k}\!\!+\!\sigma_{e}^{2}\mathbf{U}_{e,k}^{H}\mathbf{U}_{e,k}\!\!\right\}. \label{eq:MSE_cov_matrix_Eve}
\end{align}
\end{figure*}

We assume that the eavesdropper uses the well-known linear MMSE receiver\cite{Yang_MIMO_PIEE2014} for obtaining
its receive filter matrix $\mathbf{U}_{e,k}$, which is formulated as
\begin{eqnarray}\label{eq:receive_matrix_Eve}
\begin{alignedat}{1}\mathbf{U}_{e,k} & =\left(\sum_{i=1}^{K}\mathbf{H}_{ei}\mathbf{V}_{i}\mathbf{V}_{i}^{H}\mathbf{H}_{ei}^{H}+\sigma_{e}^{2}\mathbf{I}\right)^{\mathrm{-1}}\mathbf{H}_{ek}\mathbf{V}_{k}\end{alignedat},
\end{eqnarray}

We consider the following total-MSE minimization problem:
\begin{subequations}\label{eq:MT_MSE_problem}
\begin{eqnarray}
\left\{ \mathbf{V}_{k}^\star,\mathbf{U}_{k}^\star\right\}  =  \mathrm{arg}\underset{\{\mathbf{V}_{k},\mathbf{U}_{k}\}}{\mathrm{min}}\sum_{\mathit{k}=1}^{\mathit{K}}\mathrm{Tr}\left(\mathrm{\mathbf{E}_{Rx_{\mathit{k}},Tx_{\mathit{k}}}}\right), \label{eq:MT_MSE_objective} \\
\mathrm{s.t.} \quad \mathrm{\mathrm{Tr}\left(\mathrm{\mathbf{E}_{Eve,Tx_{\mathit{k}}}}\right)\geq\varepsilon_{\mathit{k}}}, \label{eq:MT_MSE_constraint} \forall k\in\left\{ 1,2,\ldots,K\right\},\\
\parallel\mathbf{V}_{k}\parallel_{F}^{2}\leq\mathbf{\mathrm{\mathit{p_{k},}}}\label{eq:MT_MSE_power_constraint}\forall k\in\left\{ 1,2,\ldots,K\right\},
\end{eqnarray}
\end{subequations}
where $\left\{ \mathbf{V}_{k}^\star,\mathbf{U}_{k}^\star\right\} $ is the solution obtained, $\mathrm{Tr}\left(\mathrm{\mathbf{E}_{Rx_{\mathit{k}},Tx_{\mathit{k}}}}\right)$
is the MSE at Receiver $k$, $\mathrm{\mathrm{Tr}\left(\mathrm{\mathbf{E}_{Eve,Tx_{\mathit{k}}}}\right)}$
is the MSE at the eavesdropper who aims for wiretapping the data symbols $\mathbf{s}_{k}$
sent from Transmitter $k$ to Receiver $k$, $\varepsilon_{\mathit{k}}$
is the threshold of the eavesdropper MSE, and $\mathbf{\mathrm{\mathit{p_{k}}}}$
is the maximum transmit power constraint imposed on Transmitter $k$.

The Lagrangian of the optimization problem
(\ref{eq:MT_MSE_problem}) is given by
\begin{align}
& \mathrm{\mathcal{L}}\left(\mathbf{V}_{k},\mathbf{U}_{k},\mu_{k},\lambda_{k}\right)\triangleq  \sum_{\mathit{k}=1}^{\mathit{K}}\mathrm{Tr}\left(\mathrm{\mathbf{E}_{Rx_{\mathit{k}},Tx_{\mathit{k}}}}\right) \nonumber \\ & \!\!+\!\!\sum_{\mathit{k}=1}^{\mathit{K}}\!\mu_{k}\!\left[\varepsilon_{\mathit{k}}\!\!-\!\!\mathrm{Tr}\left(\mathrm{\mathbf{E}_{Eve,Tx_{\mathit{k}}}}\right)\right]  \!+\!\!\sum_{\mathit{k}=1}^{\mathit{K}}\!\lambda_{k}\left[\mathrm{Tr}\left(\mathbf{V}_{k}\mathbf{V}_{k}^{H}\right)\!-\!\mathbf{\mathrm{\mathit{p_{k}}}}\right]\!,\label{eq:Lagrange_objective_function}
\end{align}
where $\mu_{k}$ and $\lambda_{k}$ are the Lagrange multipliers concerning the MSE constraint of the eavesdropper and the transmit power constraint of Transmitter $k$, respectively. In order to obtain the optimal solution of Problem (\ref{eq:MT_MSE_problem}), the following Karush-Kuhn-Tucker
(KKT) conditions have to be satisfied:  

Stationarity:
\begin{align}
\nabla_{\mathbf{V}_{k}^{*}}\mathrm{\mathcal{L}} \triangleq &   \frac{\partial\sum_{\mathit{k}=1}^{\mathit{K}}\mathrm{Tr}\left(\mathrm{\mathbf{E}_{Rx_{\mathit{k}},Tx_{\mathit{k}}}}\right)}{\partial\mathbf{V}_{k}^{*}} \nonumber \\
&  +\frac{\partial\sum_{\mathit{k}=1}^{\mathit{K}}\mu_{k}\left[\varepsilon_{\mathit{k}}-\mathrm{Tr}\left(\mathrm{\mathbf{E}_{Eve,Tx_{\mathit{k}}}}\right)\right]}{\partial\mathbf{V}_{k}^{*}} \nonumber \\ &  +\frac{\partial\sum_{\mathit{k}=1}^{\mathit{K}}\lambda_{k}\left[\mathrm{Tr}\left(\mathbf{V}_{k}\mathbf{V}_{k}^{H}\right)-\mathbf{\mathrm{\mathit{p_{k}}}}\right]}{\partial\mathbf{V}_{k}^{*}}=0,\label{eq:KKT_conditions-1}
\end{align}
\begin{align}
\nabla_{\mathbf{U}_{k}^{*}}\mathrm{\mathcal{L}}\triangleq   &  \frac{\partial\sum_{\mathit{k}=1}^{\mathit{K}}\mathrm{Tr}\left(\mathrm{\mathbf{E}_{Rx_{\mathit{k}},Tx_{\mathit{k}}}}\right)}{\partial\mathbf{U}_{k}^{*}} \nonumber \\  & +\frac{\partial\sum_{\mathit{k}=1}^{\mathit{K}}\mu_{k}\left[\varepsilon_{\mathit{k}}-\mathrm{Tr}\left(\mathrm{\mathbf{E}_{Eve,Tx_{\mathit{k}}}}\right)\right]}{\partial\mathbf{U}_{k}^{*}} \nonumber \\  & +\frac{\partial\sum_{\mathit{k}=1}^{\mathit{K}}\lambda_{k}\left[\mathrm{Tr}\left(\mathbf{V}_{k}\mathbf{V}_{k}^{H}\right)-\mathbf{\mathrm{\mathit{p_{k}}}}\right]}{\partial\mathbf{U}_{k}^{*}}=0,\label{eq:KKT_conditions-2}
\end{align}
Primal feasibility:
\begin{equation}
\mathrm{Tr}\left(\mathbf{V}_{k}\mathbf{V}_{k}^{H}\right)-\mathbf{\mathrm{\mathit{p_{k}}}}\leq0,\:\forall k\in\left\{ 1,2,\ldots,K\right\}, \label{eq:KKT_conditions-3}
\end{equation}
\begin{equation}
\varepsilon_{\mathit{k}}-\mathrm{Tr}\left(\mathrm{\mathbf{E}_{Eve,Tx_{\mathit{k}}}}\right)\leq0,\:\forall k\in\left\{ 1,2,\ldots,K\right\}, \label{eq:KKT_conditions-4}
\end{equation}
Dual feasibility:
\begin{equation}
\lambda_{k}\geq0,\:\forall k\in\left\{ 1,2,\ldots,K\right\}, \label{eq:KKT_conditions-5}
\end{equation}
\begin{equation}
\mu_{k}\geq0,\:\forall k\in\left\{ 1,2,\ldots,K\right\}, \label{eq:KKT_conditions-6}
\end{equation}
Complementary slackness:
\begin{equation}
\lambda_{k}\left[\mathrm{Tr}\left(\mathbf{V}_{k}\mathbf{V}_{k}^{H}\right)-\mathbf{\mathrm{\mathit{p_{k}}}}\right]=0,\:\forall k\in\left\{ 1,2,\ldots,K\right\}, \label{eq:KKT_conditions-7}
\end{equation}
\begin{equation}
\mu_{k}\left[\varepsilon_{\mathit{k}}-\mathrm{Tr}\left(\mathrm{\mathbf{E}_{Eve,Tx_{\mathit{k}}}}\right)\right]=0,\:\forall k\in\left\{ 1,2,\ldots,K\right\}. \label{eq:KKT_conditions-8}
\end{equation}

By substituting (\ref{eq:MSE_cov_matrix_receiver_k}) as well as (\ref{eq:MSE_cov_matrix_Eve})
into (\ref{eq:KKT_conditions-1}) and by exploiting the matrix derivative
\cite{Matrix_Cookbook,Lutkepohl_Wiley1996}, we have:
\begin{align}
\nabla_{\mathbf{V}_{k}^{*}}\mathrm{\mathcal{L}}= & \sum_{\mathit{k}=1}^{\mathit{K}}\left(\mathrm{\mathbf{H}_{\mathit{lk}}^{\mathit{H}}}\mathbf{U}_{l}\mathbf{U}_{l}^{H}\mathbf{H}_{\mathit{lk}}\right)\mathbf{V}_{k}-\mathrm{\mathbf{H}_{\mathit{kk}}^{\mathit{H}}}\mathbf{U}_{k}+\lambda_{k}\mathbf{V}_{k} \nonumber \\ & -\mu_{k}\left(\mathrm{\mathbf{H}_{\mathit{ek}}^{\mathit{H}}}\mathbf{U}_{e,k}\mathbf{U}_{e,k}^{H}\mathbf{H}_{e\mathit{k}}\mathbf{V}_{k}-\mathrm{\mathbf{H}_{\mathit{ek}}^{\mathit{H}}}\mathbf{U}_{e,k}\right)\nonumber \\
= & \!\!\left[\sum_{\mathit{k}=1}^{\mathit{K}}\!\!\left(\mathrm{\mathbf{H}_{\mathit{lk}}^{\mathit{H}}}\mathbf{U}_{l}\mathbf{U}_{l}^{H}\mathbf{H}_{\mathit{lk}}\!\right)\!\!-\!\!\mu_{k}\mathrm{\mathbf{H}_{\mathit{ek}}^{\mathit{H}}}\mathbf{U}_{e,k}\mathbf{U}_{e,k}^{H}\mathbf{H}_{e\mathit{k}} \!\!+\!\!\lambda_{k}\mathbf{I}\!\right]\!\!\mathbf{V}_{k} \nonumber \\
& -\mathrm{\mathbf{H}_{\mathit{kk}}^{\mathit{H}}}\mathbf{U}_{k}-\mu_{k}\mathrm{\mathbf{H}_{\mathit{ek}}^{\mathit{H}}}\mathbf{U}_{e,k}=0.\label{eq:derivative_V}
\end{align}

Similarly, (\ref{eq:KKT_conditions-2}) can be simplified as:
\begin{equation}
\nabla_{\mathbf{U}_{k}^{*}}\mathrm{\mathcal{L}}=\left(\sum_{\mathit{k}=1}^{\mathit{K}}\mathbf{H}_{\mathit{kl}}\mathbf{V}_{l}\mathbf{V}_{l}^{H}\mathbf{H}_{\mathit{kl}}^{\mathit{H}}+\sigma_{k}^{2}\mathbf{I}\right)\mathbf{U}_{k}-\mathbf{H}_{k\mathit{k}}\mathbf{V}_{k}=0.\label{eq:derivative_U}
\end{equation}

Finally, according to the above derivations of the first-order optimality
conditions with respect to each $\left\{ \mathbf{V}_{k},\mathbf{U}_{k}\right\} ,\: k\in\left\{ 1,2,\ldots,K\right\} $, the optimum solution of Problem (\ref{eq:MT_MSE_problem}) has to satisfy \eqref{eq:expression_V}
\begin{figure*}
\begin{eqnarray}
\mathbf{V}_{k}  =  \left[\sum_{\mathit{k}=1}^{\mathit{K}}\left(\mathrm{\mathbf{H}_{\mathit{lk}}^{\mathit{H}}}\mathbf{U}_{l}\mathbf{U}_{l}^{H}\mathbf{H}_{\mathit{lk}}\right)-\mu_{k}\mathrm{\mathbf{H}_{\mathit{ek}}^{\mathit{H}}}\mathbf{U}_{e,k}\mathbf{U}_{e,k}^{H}\mathbf{H}_{e\mathit{k}} +\lambda_{k}\mathbf{I}\right]^{-1}  \left(\mathrm{\mathbf{H}_{\mathit{kk}}^{\mathit{H}}}\mathbf{U}_{k}-\mu_{k}\mathrm{\mathbf{H}_{\mathit{ek}}^{\mathit{H}}}\mathbf{U}_{e,k}\right)\label{eq:expression_V}
\end{eqnarray}
\end{figure*}
and
\begin{eqnarray}
\mathbf{U}_{k} & = & \left(\sum_{\mathit{k}=1}^{\mathit{K}}\mathbf{H}_{\mathit{kl}}\mathbf{V}_{l}\mathbf{V}_{l}^{H}\mathbf{H}_{\mathit{kl}}^{\mathit{H}}+\sigma_{k}^{2}\mathbf{I}\right)^{-1}\mathbf{H}_{k\mathit{k}}\mathbf{V}_{k},\label{eq:expression_U}
\end{eqnarray}
where $\lambda_{k}\geq0$ and $\mu_{k}\geq0,\: k\in\left\{ 1,2,\ldots,K\right\} $
should be chosen so that the \textit{complementary slackness} conditions of (\ref{eq:KKT_conditions-7})
and (\ref{eq:KKT_conditions-8}) are satisfied.
By observing (\ref{eq:expression_V}) and (\ref{eq:expression_U}), we find that the optimal TPC and receive filter matrices $\left\{ \mathbf{V}_{k},\mathbf{U}_{k}\right\} $
depend on each other. More explicitly, each TPC matrix $\mathbf{V}_{k}$ depends
on all the receive filter matrices $\left\{ \mathbf{U}_{k}\right\} $,
while each receive filter matrix $\mathbf{U}_{k}$ depends on all
the TPC matrices $\left\{ \mathbf{V}_{k}\right\} $ as well. Hence,
it remains an open challenge to obtain the closed-form expressions for each TPC
matrix $\mathbf{V}_{k}$ and receive filter matrix $\mathbf{U}_{k}$. In what follows, a feasible approach, namely the MT-MSE algorithm in which $\mathbf{V}_{k}$
and $\mathbf{U}_{k}$ can be iteratively calculated, is proposed for determining $\left\{ \mathbf{V}_{k},\mathbf{U}_{k}\right\}$. For clarity, the MT-MSE algorithm proposed for secure communications in the MIMO interference channel is summarized in Algorithm~1.

Here, we assume that the global CSI is available at all transmitters, which is feasible for both the frequency-division duplex (FDD) and the time-division duplex (TDD) based IA systems\footnote{Obtaining the CSI of the unintended receiver (e.g. the eavesdropper) requires careful attention. Note that typically an eavesdropper has to communicate with the legitimate transmitter, e.g. a base station (BS), at some time instance to obtain certain \textit{a priori} information in support of its subsequent malicious signal processing. For example,  the eavesdropper has to estimate its own CSI ${\bf H}_{ei}$, as shown in (10), to calculate its own receive filter matrix ${\bf U}_{e,k}$.   In other words, an eavesdropper may also be a legitimate terminal of a network during a particular observation period different from the current one\cite{Fakoorian_Confidential_Interference_Channel_TIFS2011}. As a result, the eavesdropper can be regarded as a so-called ``active eavesdropper'', which also occasionally transmit feedback or pilot signals rather than just passively overhearing the signals of legitimate transmitters. Therefore, the techniques introduced in\cite{Heath_2013:IA_magazine} can be used for obtaining the eavesdropper's CSI. More details about the availability of the eavesdropper's CSI are also discussed in \cite{Koyluoglu_Secrecy_IA_TIT2011, Fakoorian_Confidential_Interference_Channel_TIFS2011}. Furthermore, for fast time-varying fading channels where accurate CSI is difficult to obtain, the eavesdropper's outdated CSI and statistical CSI may still be attainable at the legitimate transmitters for optimizing the TPC matrix.}, as shown in\cite{Heath_2013:IA_magazine}. Consequently, each transmitter is capable of calculating its TPC matrix. To elaborate a little further, Transmitter $k$ starts with a random TPC matrix, which is generated according to $\mathcal{CN}(0,1)$ and then normalized in order to satisfy the transmit power constraint. At the receiver $k$ and the eavesdropper, $\mathbf{U}_{k}$ and $\mathbf{U}_{e,k}$ are respectively calculated according to the MMSE criterion of (\ref{eq:receive_matrix_Eve}). At the transmitter $k$, we consider the challenging situation where the eavesdropper is capable of calculating its receive filter matrix $\mathbf{U}_{e,k}$ according to the MMSE criterion of (\ref{eq:receive_matrix_Eve}). The major part of our MT-MSE algorithm, which describes how to calculate the TPC matrix, is summarized as follows.
\begin{enumerate}
\item
Firstly, the Lagrange multiplier $\lambda_{k}$ is calculated according to the transmit power constraint characterized by (\ref{eq:MT_MSE_power_constraint}) and (\ref{eq:KKT_conditions-7}). Then, we use the above $\lambda_{k}$ to calculate and update the intermediate result of the TPC matrix, as shown in (\ref{eq:intermediate_result_V}).
\item
Secondly, the Lagrange multiplier $\mu_{k}$ is calculated according to the intermediate result of the TPC matrix obtained above and to the MSE constraint, which is characterized by (\ref{eq:MT_MSE_constraint}) and (\ref{eq:KKT_conditions-8}).
\item
Finally, we obtain the TPC matrix by substituting the above $\lambda_{k}$ and $\mu_{k}$ into (\ref{eq:updated_V}). The details of our MT-MSE algorithm are presented in Algorithm 1.
\end{enumerate}
\begin{figure*}
\onecolumn
\begin{algorithm}[H]
\small
\protect\caption{\small The Proposed Secure Iterative Distributed MT-MSE Algorithm (Part I)}

\begin{enumerate}
\item[\textbf{1}.] Initialization: set the iteration counter to $n=0$, and start with an arbitrary TPC matrix $\mathbf{V}_{k}^{\left(n\right)} = \mathbf{V}_{k}^{\left(0\right)}$, $k\in\left\{ 1,2,\ldots,K\right\} $.
\item[\textbf{2}.] Begin the iteration: calculate and update $\mathbf{U}_{k}^{\left(n+1\right)}$
and $\mathbf{U}_{e,k}^{\left(n+1\right)}$ according to the MMSE receiver criterion of (\ref{eq:receive_matrix_Eve}). Then we have
\begin{align}
\mathbf{U}_{k}^{\left(n+1\right)} & =\left[\sum_{\mathit{l}=1}^{\mathit{K}}\mathbf{H}_{\mathit{kl}}\mathbf{V}_{l}^{\left(n\right)}\left(\mathbf{V}_{l}^{\left(n\right)}\right)^{H}\mathbf{H}_{\mathit{kl}}^{\mathit{H}}+\sigma_{k}^{2}\mathbf{I}\right]^{-1}\mathbf{H}_{k\mathit{k}}\mathbf{V}_{k}^{\left(n\right)},\: k\in\left\{ 1,2,\ldots,K\right\} ,\label{eq:U_k}
\end{align}
\begin{align}
\mathbf{U}_{e,k}^{\left(n+1\right)} & =\left[\sum_{\mathit{i}=1}^{\mathit{K}}\mathbf{H}_{\mathit{ei}}\mathbf{V}_{i}^{\left(n\right)}\left(\mathbf{V}_{i}^{\left(n\right)}\right)^{H}\mathbf{H}_{\mathit{ei}}^{\mathit{H}}+\sigma_{e}^{2}\mathbf{I}\right]^{-1}\mathbf{H}_{e\mathit{k}}\mathbf{V}_{k}^{\left(n\right)},\: k\in\left\{ 1,2,\ldots,K\right\} .\label{eq:U_ek}
\end{align}

\item[\textbf{3}.] Calculate and update $\hat{\lambda}_{k}^{\left(n+1\right)}$:

\begin{enumerate}
\item[\textbf{i}.] \textbf{if }$n=0$, use the given $\mu_{k}^{\left(0\right)}$, \textbf{else} use $\mu_{k}^{\left(n\right)}$ that has been calculated.
\item[\textbf{ii}.] Calculate $\hat{\lambda}_{k}^{\left(n+1\right)}$ according to the equation
$\mathrm{Tr}\left[\mathbf{\hat{V}}_{k}^{\left(n+1\right)}\left(\mathbf{\hat{V}}_{k}^{\left(n+1\right)}\right)^{H}\right]=\mathbf{\mathrm{\mathit{p_{k}}}},\: k\in\left\{ 1,2,\ldots,K\right\} $
(the solution of $\hat{\lambda}_{k}^{\left(n+1\right)}$ is detailed
in Appendix \ref{Appendix A}), where
\begin{align}
\mathbf{\hat{V}}_{k}^{\left(n+1\right)}\left(\hat{\lambda}_{k}^{\left(n+1\right)}\right)  = & \left\{ \sum_{\mathit{k}=1}^{\mathit{K}}\left[\mathrm{\mathbf{H}_{\mathit{lk}}^{\mathit{H}}}\mathbf{U}_{l}^{\left(n+1\right)}\left(\mathbf{U}_{l}^{\left(n+1\right)}\right)^{H}\mathbf{H}_{\mathit{lk}}\right]-\mu_{k}^{\left(n\right)}\mathrm{\mathbf{H}_{\mathit{ek}}^{\mathit{H}}}\mathbf{U}_{e,k}^{\left(n+1\right)}\left(\mathbf{U}_{e,k}^{\left(n+1\right)}\right)^{H}\mathbf{H}_{e\mathit{k}} + \hat{\lambda}_{k}^{\left(n+1\right)}\mathbf{I}\right\} ^{-1} \nonumber \\
   & \times \left(\mathrm{\mathbf{H}_{\mathit{kk}}^{\mathit{H}}}\mathbf{U}_{k}^{\left(n+1\right)}-\mu_{k}^{\left(n\right)}\mathrm{\mathbf{H}_{\mathit{ek}}^{\mathit{H}}}\mathbf{U}_{e,k}^{\left(n+1\right)}\right),\: k\in\left\{ 1,2,\ldots,K\right\} .\label{eq:V^hat}
\end{align}

\end{enumerate}
\item[\textbf{4}.] Update $\lambda_{k}^{\left(n+1\right)}$ according to  $\lambda_{k}^{\left(n+1\right)}=\mathrm{max}\left(\hat{\lambda}_{k}^{\left(n+1\right)},0\right),\: k\in\left\{ 1,2,\ldots,K\right\} .$
\item[\textbf{5}.] Calculate and update the intermediate result of the TPC matrix $\mathbf{\bar{V}}_{k}^{\left(n+1\right)}$ according to
\begin{align}
\mathbf{\bar{V}}_{k}^{\left(n+1\right)} & =\left\{ \sum_{\mathit{l}=1}^{\mathit{K}}\left[\mathrm{\mathbf{H}_{\mathit{lk}}^{\mathit{H}}}\mathbf{U}_{l}^{\left(n+1\right)}\left(\mathbf{U}_{l}^{\left(n+1\right)}\right)^{H}\mathbf{H}_{\mathit{lk}}\right]-\mu_{k}^{\left(n\right)}\mathrm{\mathbf{H}_{\mathit{ek}}^{\mathit{H}}}\mathbf{U}_{e,k}^{\left(n+1\right)}\left(\mathbf{U}_{e,k}^{\left(n+1\right)}\right)^{H}\mathbf{H}_{e\mathit{k}}\right.\nonumber \\
 & \quad\left.+\lambda_{k}^{\left(n+1\right)}\mathbf{I}\right\} ^{-1}\left(\mathrm{\mathbf{H}_{\mathit{kk}}^{\mathit{H}}}\mathbf{U}_{k}^{\left(n+1\right)}-\mu_{k}^{\left(n\right)}\mathrm{\mathbf{H}_{\mathit{ek}}^{\mathit{H}}}\mathbf{U}_{e,k}^{\left(n+1\right)}\right),\: k\in\left\{ 1,2,\ldots,K\right\} .\label{eq:intermediate_result_V}
\end{align}
\\
\end{enumerate}

\end{algorithm}
\twocolumn
\end{figure*}

\begin{figure*}
\onecolumn
\begin{algorithm}[H]
\small
\protect\caption*{\small \textbf{Algorithm 1} The Proposed Secure Iterative Distributed MT-MSE Algorithm (Part II)}

\begin{enumerate}

\item[\textbf{6}.] Calculate and update $\mu_{k}^{\left(n+1\right)}$ and the final result of the TPC matrix $\mathbf{V}_{k}^{\left(n+1\right)}$:

\begin{enumerate}
\item[\textbf{i}.] Set\textbf{ $k=1$}.\textbf{ }
\item[\textbf{ii}.] \textbf{if }$k=1$, solve for $\widetilde{\mu}_{k}^{\left(n+1\right)}$
relying on $\mathrm{Tr}\left\{ \mathrm{\mathbf{E}_{Eve,Tx_{\mathit{k}}}\left[\mathbf{\widetilde{V}}_{\mathit{k}}^{\left(\mathit{n}+1\right)}\left(\widetilde{\mu}_{\mathit{k}}^{\left(\mathit{n}+1\right)}\right)\right]}\right\} =\varepsilon_{\mathit{k}}$.
\begin{align}
\mathrm{\Rightarrow} & \mathrm{Tr}\left\{ \left(\mathbf{U}_{e,k}^{\left(n+1\right)}\right)^{H}\left[\sum_{i=2}^{K}\mathbf{H}_{ei}\mathbf{\bar{V}}_{i}^{\left(n+1\right)}\left(\mathbf{\bar{V}}_{i}^{\left(n+1\right)}\right)^{H}\mathbf{H}_{ei}^{H}\right]\mathbf{U}_{e,k}^{\left(n+1\right)}\right.\nonumber \\
 & +\left(\mathbf{U}_{e,k}^{\left(n+1\right)}\right)^{H}\mathbf{H}_{ei}\mathbf{\widetilde{V}}_{k}^{\left(n+1\right)}\left(\mathbf{\widetilde{V}}_{k}^{\left(n+1\right)}\right)^{H}\mathbf{H}_{ei}^{H}\mathbf{U}_{e,k}^{\left(n+1\right)}-\left(\mathbf{U}_{e,k}^{\left(n+1\right)}\right)^{H}\mathbf{H}_{ek}\mathbf{\widetilde{V}}_{k}^{\left(n+1\right)}\nonumber \\
 & \left.-\left(\mathbf{\widetilde{V}}_{k}^{\left(n+1\right)}\right)^{H}\mathbf{H}_{ek}^{H}\mathbf{U}_{e,k}^{\left(n+1\right)}+\sigma_{e}^{2}\left(\mathbf{U}_{e,k}^{\left(n+1\right)}\right)^{H}\mathbf{U}_{e,k}^{\left(n+1\right)}+\mathbf{I}\right\} =\varepsilon_{\mathit{k}}.\label{eq:calculate_mu_from_MSE_Eve-1}
\end{align}
\textbf{else}\\
use the calculated $\mathbf{V}_{i}^{\left(n+1\right)}$ that is the
matrix function of $\mu_{i}^{\left(n+1\right)},\: i\in\left\{ 1,2,\ldots,k-1\right\} $, and solve for $\widetilde{\mu}_{k}^{\left(n+1\right)}$ from $\mathrm{Tr}\left\{ \mathrm{\mathbf{E}_{Eve,Tx_{\mathit{k}}}\left[\mathbf{\widetilde{V}}_{\mathit{k}}^{\left(\mathit{n}+1\right)}\left(\widetilde{\mu}_{\mathit{k}}^{\left(\mathit{n}+1\right)}\right)\right]}\right\} =\varepsilon_{\mathit{k}}$.
\begin{align}
\mathrm{\Rightarrow} & \mathrm{Tr}\left\{ \left(\mathbf{U}_{e,k}^{\left(n+1\right)}\right)^{H}\left[\sum_{i=1}^{k-1}\mathbf{H}_{ei}\mathbf{V}_{i}^{\left(n+1\right)}\left(\mathbf{V}_{i}^{\left(n+1\right)}\right)^{H}\mathbf{H}_{ei}^{H}\right]\mathbf{U}_{e,k}^{\left(n+1\right)}\right.\nonumber \\
 & +\left(\mathbf{U}_{e,k}^{\left(n+1\right)}\right)^{H}\left[\sum_{i=k+1}^{K}\mathbf{H}_{ei}\mathbf{\bar{V}}_{i}^{\left(n+1\right)}\left(\mathbf{\bar{V}}_{i}^{\left(n+1\right)}\right)^{H}\mathbf{H}_{ei}^{H}\right]\mathbf{U}_{e,k}^{\left(n+1\right)}\nonumber \\
 & +\left(\mathbf{U}_{e,k}^{\left(n+1\right)}\right)^{H}\mathbf{H}_{ei}\mathbf{\widetilde{V}}_{k}^{\left(n+1\right)}\left(\mathbf{\widetilde{V}}_{k}^{\left(n+1\right)}\right)^{H}\mathbf{H}_{ei}^{H}\mathbf{U}_{e,k}^{\left(n+1\right)}-\left(\mathbf{U}_{e,k}^{\left(n+1\right)}\right)^{H}\mathbf{H}_{ek}\mathbf{\widetilde{V}}_{k}^{\left(n+1\right)}\nonumber \\
 & \left.-\left(\mathbf{\widetilde{V}}_{k}^{\left(n+1\right)}\right)^{H}\mathbf{H}_{ek}^{H}\mathbf{U}_{e,k}^{\left(n+1\right)}+\sigma_{e}^{2}\left(\mathbf{U}_{e,k}^{\left(n+1\right)}\right)^{H}\mathbf{U}_{e,k}^{\left(n+1\right)}+\mathbf{I}\right\} =\varepsilon_{\mathit{k}}.\label{eq:calculate_mu_from_MSE_Eve-2}
\end{align}
\textbf{end}\\
According to (\ref{eq:calculate_mu_from_MSE_Eve-1}) or (\ref{eq:calculate_mu_from_MSE_Eve-2}), we obtain $\widetilde{\mu}_{k}^{\left(n+1\right)}=\left\{ \widetilde{\mu}_{k}^{\left(n+1\right)}\mid\mathrm{Tr}\left\{ \mathrm{\mathbf{E}_{Eve,TX_{\mathit{k}}}\left[\mathbf{\widetilde{V}}_{\mathit{k}}^{\left(\mathit{n}+1\right)}\left(\widetilde{\mu}_{\mathit{k}}^{\left(\mathit{n}+1\right)}\right)\right]}\right\} =\varepsilon_{\mathit{k}}\right\} $,
$k\in\left\{ 1,2,\ldots,K\right\} $ (the solution of $\widetilde{\mu}_{k}^{\left(n+1\right)}$
is detailed in Appendix \ref{Appendix B}), where
\begin{align}
\mathbf{\mathbf{\widetilde{V}}}_{k}^{\left(n+1\right)} & =\left\{ \sum_{\mathit{l}=1}^{\mathit{K}}\left[\mathrm{\mathbf{H}_{\mathit{lk}}^{\mathit{H}}}\mathbf{U}_{l}^{\left(n+1\right)}\left(\mathbf{U}_{l}^{\left(n+1\right)}\right)^{H}\mathbf{H}_{\mathit{lk}}\right]-\widetilde{\mu}_{k}^{\left(n+1\right)}\mathrm{\mathbf{H}_{\mathit{ek}}^{\mathit{H}}}\mathbf{U}_{e,k}^{\left(n+1\right)}\left(\mathbf{U}_{e,k}^{\left(n+1\right)}\right)^{H}\mathbf{H}_{e\mathit{k}}\right.\nonumber \\
 & \quad\left.+\lambda_{k}^{\left(n+1\right)}\mathbf{I}\right\} ^{-1}\left(\mathrm{\mathbf{H}_{\mathit{kk}}^{\mathit{H}}}\mathbf{U}_{k}^{\left(n+1\right)}-\widetilde{\mu}_{k}^{\left(n+1\right)}\mathrm{\mathbf{H}_{\mathit{ek}}^{\mathit{H}}}\mathbf{U}_{e,k}^{\left(n+1\right)}\right),\: k\in\left\{ 1,2,\ldots,K\right\} .\label{eq:V^tilde}
\end{align}
\item[\textbf{iii}.] Update $\mu_{k}^{\left(n+1\right)}$ according to $\mu_{k}^{\left(n+1\right)}=\mathrm{max}\left(\widetilde{\mu}_{k}^{\left(n+1\right)},0\right),\: k\in\left\{ 1,2,\ldots,K\right\} $.
\item[\textbf{iv}.] According to the above\textbf{ $\mu_{k}^{\left(n+1\right)}$}, calculate
and update the final result of the TPC matrix \textbf{$\mathrm{\mathbf{V}}_{k}^{\left(n+1\right)}$} according to
\begin{align}
\mathbf{\mathbf{V}}_{k}^{\left(n+1\right)} & =\left\{ \sum_{\mathit{l}=1}^{\mathit{K}}\left[\mathrm{\mathbf{H}_{\mathit{lk}}^{\mathit{H}}}\mathbf{U}_{l}^{\left(n+1\right)}\left(\mathbf{U}_{l}^{\left(n+1\right)}\right)^{H}\mathbf{H}_{\mathit{lk}}\right]-\mu_{k}^{\left(n+1\right)}\mathrm{\mathbf{H}_{\mathit{ek}}^{\mathit{H}}}\mathbf{U}_{e,k}^{\left(n+1\right)}\left(\mathbf{U}_{e,k}^{\left(n+1\right)}\right)^{H}\mathbf{H}_{e\mathit{k}}\right.\nonumber \\
 & \quad\left.+\lambda_{k}^{\left(n+1\right)}\mathbf{I}\right\} ^{-1}\left(\mathrm{\mathbf{H}_{\mathit{kk}}^{\mathit{H}}}\mathbf{U}_{k}^{\left(n+1\right)}-\mu_{k}^{\left(n+1\right)}\mathrm{\mathbf{H}_{\mathit{ek}}^{\mathit{H}}}\mathbf{U}_{e,k}^{\left(n+1\right)}\right),\: k\in\left\{ 1,2,\ldots,K\right\} .\label{eq:updated_V}
\end{align}

\item[\textbf{v}.] \textbf{$k\leftarrow k+1$}. 
Repeat step \textbf{ii},
\textbf{iii}, \textbf{iv} and \textbf{v} until\textbf{ $k>K$}.
\end{enumerate}
\item[\textbf{7}.] Repeat Step 2 through 6 until $\mathbf{\mathbf{V}}_{k}^{\left(n+1\right)}$
satisfies the transmit power constraint and the algorithm has converged.
\end{enumerate}
\end{algorithm}
\twocolumn
\end{figure*}

\section{Convergence of the Proposed Secure MT-MSE Algorithm}
\label{sub:Convergence}
In this section, we demonstrate that the proposed MT-MSE algorithm is guaranteed to converge.
Since the TPC matrices $\left\{ \mathbf{V}_{k}\right\} $ and receive
filter matrices $\left\{ \mathbf{U}_{k}\right\} $ are iteratively
calculated at each iteration of the proposed MT-MSE algorithm, the value of the total-MSE objective function $\sum_{\mathit{k}=1}^{\mathit{K}}\mathrm{Tr}\left(\mathrm{\mathbf{E}_{Rx_{\mathit{k}},Tx_{\mathit{k}}}}\right)$ is monotonically
reduced. The strict proof of the convergence is as follows.

For given values of the TPC matrices $\left\{ \mathbf{V}_{k}\right\} $,
the total-MSE minimization problem (\ref{eq:MT_MSE_problem})
can be reformulated as:
\begin{subequations}\label{eq:MT_MSE_problem_reformulated}
\begin{eqnarray}
\left\{ \mathbf{U}_{k}^\star\right\} =\mathrm{arg}\; \underset{\mathbf{U}_{k}} {\mathrm{min}}\sum_{\mathit{k}=1}^{\mathit{K}}\mathrm{Tr}\left(\mathrm{\mathbf{E}_{Rx_{\mathit{k}},Tx_{\mathit{k}}}}\right), \label{eq:MT_MSE_problem_reformulated_objective}\\
\mathrm{s.t.}\;\mathrm{\mathrm{Tr}\left(\mathrm{\mathbf{E}_{Eve,Tx_{\mathit{k}}}}\right)\geq\varepsilon_{\mathit{k}}},\;\forall k\in\left\{ 1,2,\ldots,K\right\} .\label{eq:MT_MSE_function_for_the_given_V}
\end{eqnarray}
\end{subequations}
The solution to this constrained optimization problem (\ref{eq:MT_MSE_problem_reformulated})
is given by (\ref{eq:expression_U}). Hence, we have \eqref{eq:proof_convergency_1}:
\begin{eqnarray}
 \sum_{\mathit{k}=1}^{\mathit{K}}\mathrm{Tr}\left[\mathrm{\mathbf{E}_{Rx_{\mathit{k}},Tx_{\mathit{k}}}}\left(\mathbf{V}_{k}^{\left(n\right)},\mathbf{U}_{k}^{\left(n+1\right)}\right)\right]
 \leq\nonumber\\
 \sum_{\mathit{k}=1}^{\mathit{K}}\mathrm{Tr}\left[\mathrm{\mathbf{E}_{Rx_{\mathit{k}},Tx_{\mathit{k}}}}\left(\mathbf{V}_{k}^{\left(n\right)},\mathbf{U}_{k}^{\left(n\right)}\right)\right].\label{eq:proof_convergency_1}
\end{eqnarray}

Likewise, given the values of the receive filter matrices $\left\{ \mathbf{U}_{k}\right\} $,
the total-MSE minimization problem (\ref{eq:MT_MSE_problem})
can be rewritten as:
\begin{subequations}\label{eq:MT_MSE_problem_reformulated_2}
\begin{eqnarray}
\left\{ \mathbf{V}_{k}^\star\right\} =\mathrm{arg}\;\underset{\mathbf{V}_{k}} {\mathrm{min}}\sum_{\mathit{k}=1}^{\mathit{K}}\mathrm{Tr}\left(\mathrm{\mathbf{E}_{Rx_{\mathit{k}},Tx_{\mathit{k}}}}\right), \label{eq:MT_MSE_problem_objective_2} \\
\mathrm{s.t.}\;\mathrm{\mathrm{Tr}\left(\mathrm{\mathbf{E}_{Eve,Tx_{\mathit{k}}}}\right)\geq\varepsilon_{\mathit{k}}},\;\forall k\in\left\{ 1,2,\ldots,K\right\} , \\
\parallel\mathbf{V}_{k}\parallel_{F}^{2}\leq\mathbf{\mathrm{\mathit{p_{k},\;\forall k\in\left\{ \mathrm{1,2,\ldots,}K\right\} }}}.\label{eq:MT_MSE_function_for_the_given_U}
\end{eqnarray}
\end{subequations}
The solution to this constrained optimization problem (\ref{eq:MT_MSE_problem_reformulated_2})
is given by (\ref{eq:expression_V}). Similarly, we have \eqref{eq:proof_convergency_2}:
\begin{eqnarray}
\sum_{\mathit{k}=1}^{\mathit{K}}\mathrm{Tr}\left[\mathrm{\mathbf{E}_{Rx_{\mathit{k}},Tx_{\mathit{k}}}}\left(\mathbf{V}_{k}^{\left(n+1\right)},\mathbf{U}_{k}^{\left(n+1\right)}\right)\right]\leq\nonumber\\
\sum_{\mathit{k}=1}^{\mathit{K}}\mathrm{Tr}\left[\mathrm{\mathbf{E}_{Rx_{\mathit{k}},Tx_{\mathit{k}}}}\left(\mathbf{V}_{k}^{\left(n\right)},\mathbf{U}_{k}^{\left(n+1\right)}\right)\right].\label{eq:proof_convergency_2}
\end{eqnarray}

\begin{figure}
\begin{centering}
\includegraphics[width=3.5in]{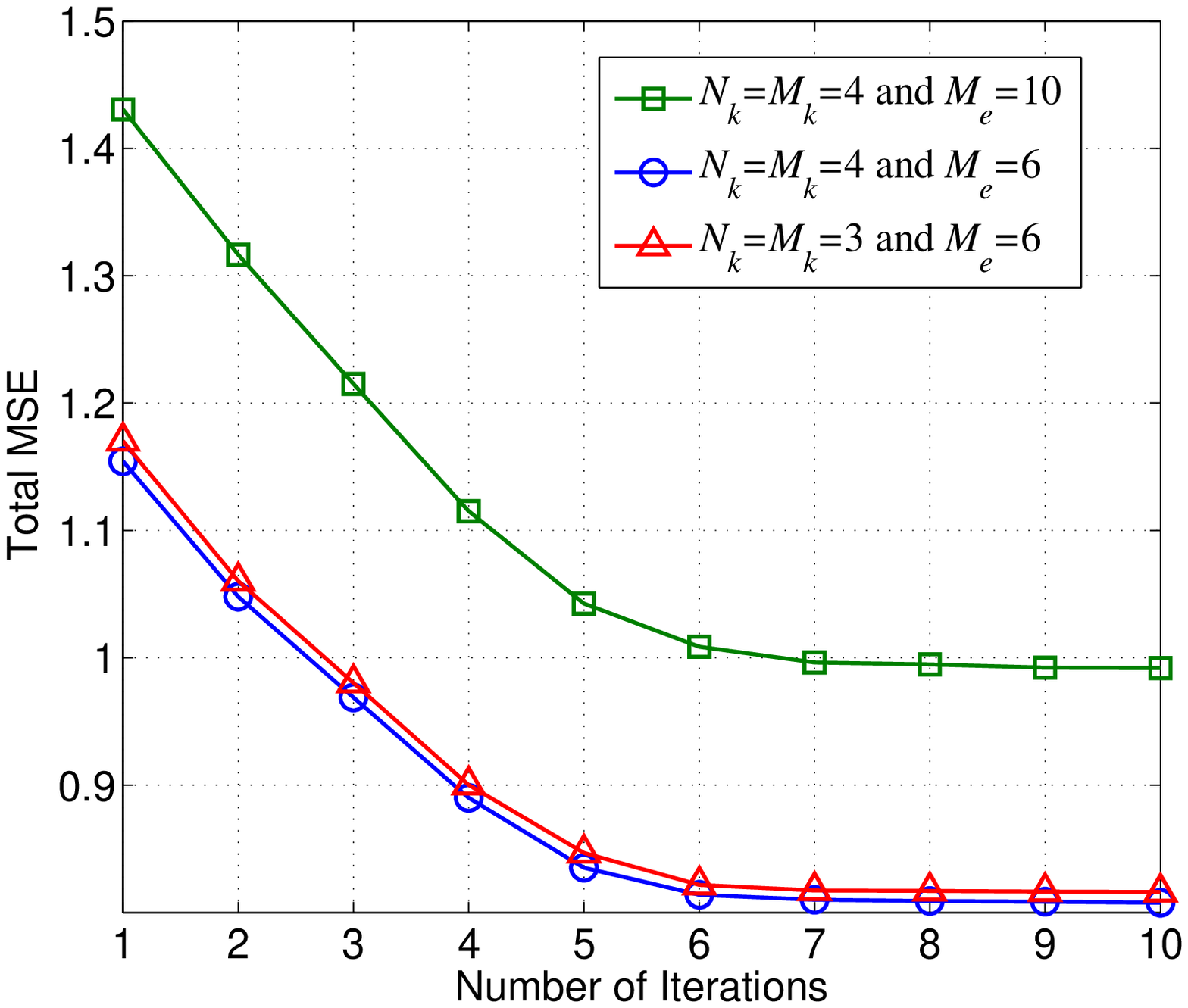}
\par\end{centering}
\centering{}\protect\caption{Convergence example of the MT-MSE algorithm. We have $K = 3$, $\mathsf{SNR} = 25$dB and the specific SNR definition is given in Section \ref{Section:Simulation}. }
\label{fig:convergency_MSE}
\end{figure}
From the above \eqref{eq:proof_convergency_1} and \eqref{eq:proof_convergency_2}, it may be readily seen that the value of the total MSE of all receivers
is monotonically reduced after each iteration. This conclusion is also demonstrated by the numerical results of Fig.~\ref{fig:convergency_MSE}, which illustrates the convergence behavior of the MT-MSE algorithm in the MIMO interference network considered, where we have $\mathsf{SNR}=25$dB, $K = 3$, as well as diverse values of $M_k$, $N_k$ and $M_{e}$. The plot shows that the total MSE of all receivers under the three different antenna configurations considered always converges within about 6~$\sim$~7 iterations and it does indeed monotonically decrease upon increasing the number of iterations, as shown by the above analytical results. Therefore, the convergence of the MT-MSE algorithm is guaranteed. Moreover, the plot also shows that the total MSE increases upon increasing the number of eavesdropper antennas. This is because the MT-MSE algorithm limits the detection performance of the ``sophisticated/strong'' eavesdropper at the expense of increasing the total MSE of the legitimate receivers. Additionally, when the ratio of the number of antennas at each legitimate transmitter/receiver to that of the eavesdropper is increased, the total MSE of the legitimate receivers is reduced, since the relative strength of the eavesdropper erodes.

\section{Simulation Results}
\label{Section:Simulation}
In this section, we conduct a range of numerical simulations to evaluate the performance
of the proposed MT-MSE algorithm presented in Section III for the MIMO interference
channel. Here we consider a three-user MIMO interference channel, where
the number of transmit and receive antennas is identical at each of all the legitimate
transmitter-receiver pairs. All channel coefficients are assumed to be i.i.d. zero-mean unit-variance complex-valued Gaussian random variables. The transmitted power at each transmitter is the same,
i.e. we have $p_{1}=p_{2}=p_{3}$. The noise power at all receivers and the eavesdropper
is also identical. We define the signal-to-noise ratio (SNR) (in dB) as the ratio
of the received signal power at a single receiver over the noise power, i.e. we have
$\mathsf{SNR} = 10\log_{10}p_{k}/\sigma_{k}^{2}$. To demonstrate the security and effectiveness of
our MT-MSE algorithm in comparison to the traditional IA algorithm \cite{Gomadam_TIT2011,Sasaki_Secure_IA_ISAP2012}, we intentionally opt for assuming the presence of a sophisticated eavesdropper, whose number of antennas $M_{e}$ is higher than that
of each legitimate transmitter and receiver. To guarantee secure communications,
we set the threshold of the eavesdropper's MSE to $\varepsilon_{\mathit{k}}> 1$
\cite{MinyanPei_MIMO_Wiretap_with_Imperfect_CSI_TWC_2012,Reboredo_sumMSE_transmitter_power_constraints_TSP_2013},
for $k\in\left\{ 1,2,\ldots,K\right\} $. Here we evaluate the scenario
where the number of transmit and receive antennas at each legitimate
transmitter and receiver is identical, i.e. we have $N_{k}=M_{k}=M$, $k\in\left\{ 1,2,\ldots,K\right\} $.

For the above scenario, the rate (bits/second/Hertz) corresponding to the $s$th data stream of the transmitter-receiver pair $k$ is
given by
\begin{equation}
R_{k}^{\left(s\right)}=\mathrm{log}_{2}\left(1+\frac{\left\Vert \mathbf{U}_{\mathit{k}\left(s\right)}^{\mathit{H}}\mathbf{H}_{\mathit{kk}}\mathbf{V}_{\mathit{k}\left(s\right)}\right\Vert ^{2}}{Z_{k}^{\left(s\right)}}\right),k\in\left\{ \mathrm{1,2,\ldots,}K\right\}, \label{eq:rate_per_stream}
\end{equation}
where $\mathbf{V}_{k\left(s\right)}$ is the TPC vector for the $s$th data stream of the transmitter-receiver pair $k$ and the numerator of the right-hand
side is the desired signal power, while
the first, second and third terms of the denominator $Z_{k}^{\left(s\right)}$ that is given by
\begin{align}
Z_{k}^{\left(s\right)} & \!\!\!\triangleq\!\!\sum\limits _{{t=1\atop t\ne s}}^{d_{k}}\left\Vert \mathbf{U}_{\mathit{k}\left(s\right)}^{\mathit{H}}\mathbf{H}_{kk}\mathbf{V}_{k\left(t\right)}\right\Vert ^{2}\!\!\!+\!\sum\limits _{{l=1\atop l\ne k}}^{K}\sum_{t=1}^{d_{l}}\left\Vert \mathbf{U}_{\mathit{k}\left(s\right)}^{\mathit{H}}\mathbf{H}_{kl}\mathbf{V}_{l\left(t\right)}\right\Vert ^{2}\!\!\!+\!\sigma_{k}^{2}, \nonumber \\
& k\in\left\{ 1,2,\ldots,K\right\}, \label{eq:denominator_sum_rate_per_stream}
\end{align}
represent the inter-stream interference, the multiuser interference and the AWGN, respectively. Additionally, in \eqref{eq:denominator_sum_rate_per_stream} $\mathbf{V}_{l\left(t\right)}$ is the TPC
vector for the $t$th data stream of the transmitter-receiver pair $l$, and $\mathbf{U}_{k\left(s\right)}$
is the receive filter vector for the $s$th data stream of the transmitter-receiver pair $k$.
\begin{figure*}[tbp]
\begin{center}
\subfloat[Communication and secrecy rate for $M=4$, $M_{e}=6$]{\includegraphics[width=3.5in]{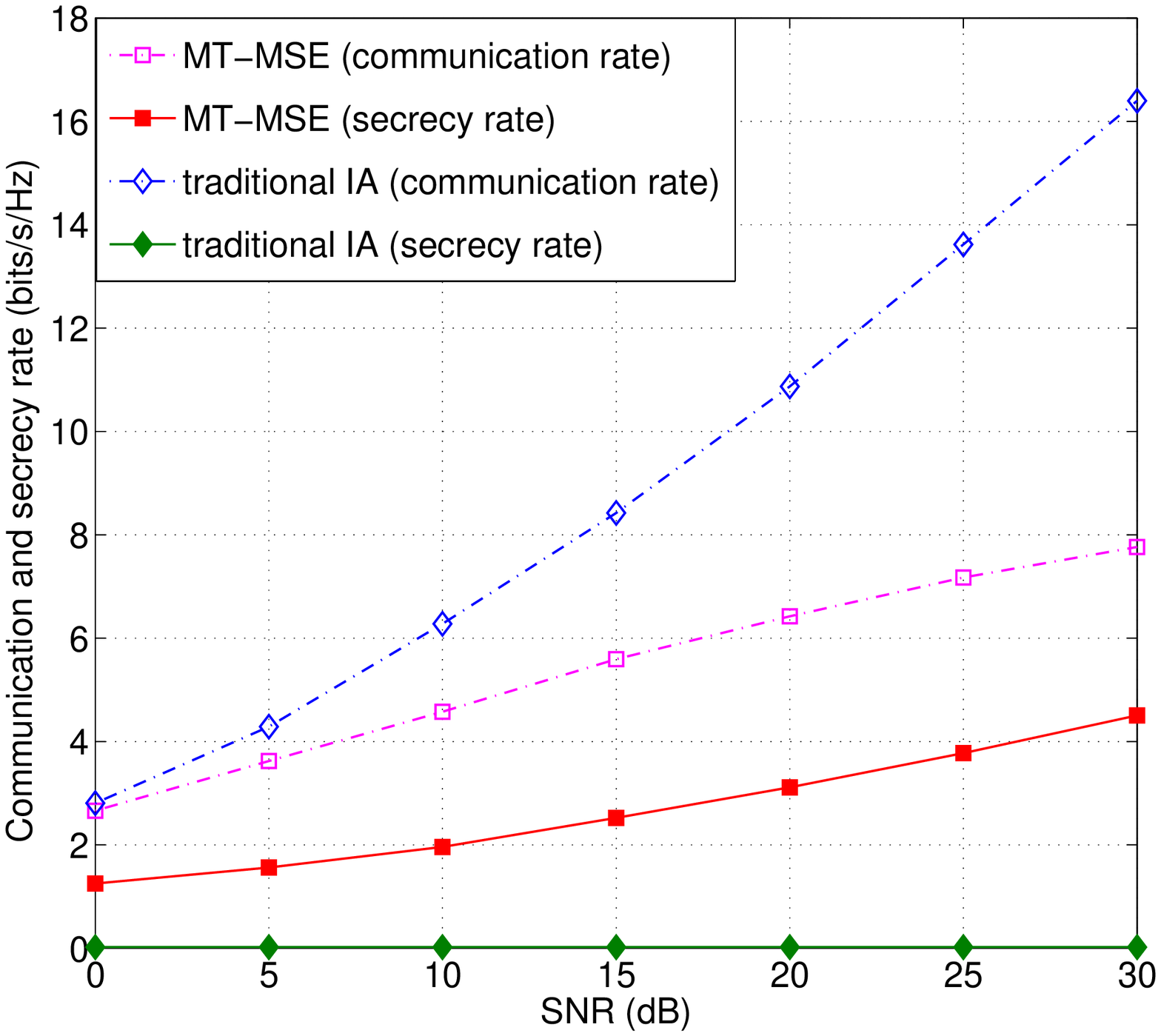}
\label{subfig:communication and secrecy rate Me=00003D6}}
\subfloat[Achievable MSE versus SNR]{\includegraphics[width=3.5in]{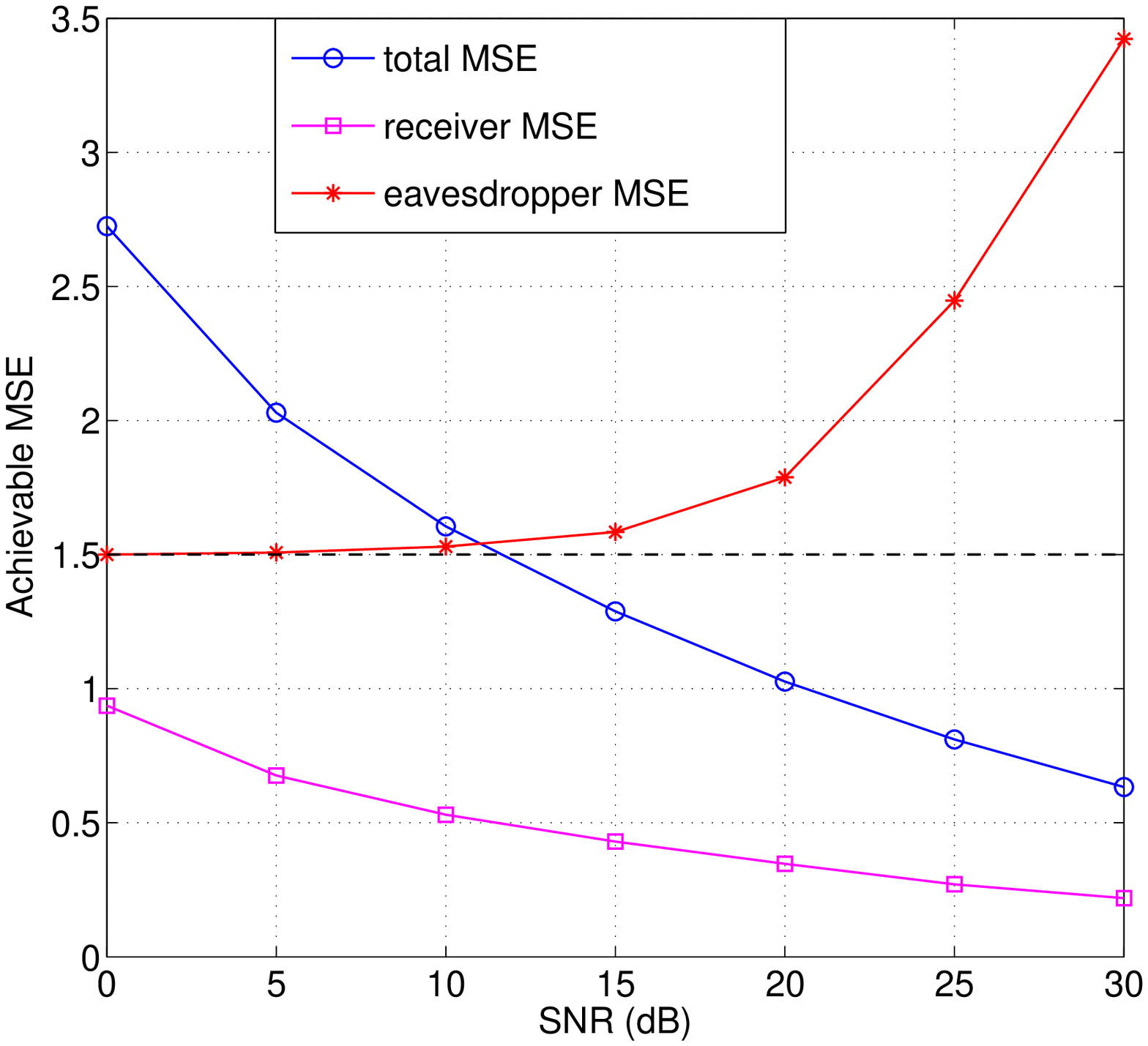}
\label{subfig:average MSE}}
\par
\end{center}
\raggedright{}\protect\caption{Communication rate, secrecy rate and MSE versus SNR performance
comparison of the proposed MT-MSE algorithm and the traditional IA
algorithm \cite{Gomadam_TIT2011,Sasaki_Secure_IA_ISAP2012} in the presence of a
``sophisticated eavesdropper'', where we have $M=4$, $M_{e}=6$ and $\varepsilon_{\mathit{k}}=1.5$.}
\label{fig:rate_MSE_Nk=00003DMk=00003D4_Me=00003D6}
\end{figure*}

Secure communications can be realized when the communication rate
between the legitimate transmitter-receiver pair is higher than that of the transmitter-eavesdropper link. The difference between these two rates is the so-called ``secrecy rate'', which is invoked in this paper for the performance evaluation of
secure communications. Explicitly, the secrecy rate of Receiver $k$ is defined as
\begin{align}
S_{\mathit{k}} & \triangleq R_{\mathrm{\mathit{k}}}-R_{\mathrm{\mathit{k},\,\mathrm{Eve}}}\nonumber \\
 & \triangleq\sum_{s=1}^{d_{k}}R_{k}^{\left(s\right)}-\sum_{s=1}^{d_{k}}\mathrm{log}_{2}\left(1+\frac{\left\Vert \mathbf{U}_{\mathit{e,k}\left(s\right)}^{\mathit{H}}\mathbf{H}_{\mathit{ek}}\mathbf{V}_{\mathit{k}\left(s\right)}\right\Vert ^{2}}{\mathbf{U}_{\mathit{e,k}\left(s\right)}^{\mathit{H}}\boldsymbol{\Upsilon}_{ek}^{\left(s\right)}\mathbf{U}_{\mathit{e,k}\left(s\right)}}\right), \nonumber \\
& k\in\left\{ \mathrm{1,2,\ldots,}K\right\}, \label{eq:secrecy_rate}
\end{align}
where we have
\begin{align}
\boldsymbol{\Upsilon}_{ek}^{\left(s\right)} & \triangleq\sum\limits _{l=1}^{d_{k}}\mathbf{H}_{\mathit{el}}\mathbf{V}_{\mathit{l}}\mathbf{V}_{\mathit{l}}^{H}\mathbf{H}_{\mathit{el}}^{H}-\mathbf{H}_{\mathit{ek}}\mathbf{V}_{\mathit{k\left(s\right)}}\mathbf{V}_{\mathit{k\left(s\right)}}^{H}\mathbf{H}_{\mathit{ek}}^{H}+\sigma_{k}^{2}\mathbf{I},\nonumber \\
& k\in\left\{ \mathrm{1,2,\ldots,}K\right\},\label{eq:denominator_term_Eve}
\end{align}
and $\mathbf{U}_{e,k\left(s\right)}$ is the eavesdropper's wiretap receive filter vector for the $s$th data stream of the transmitter-receiver pair $k$. 

Fig. \ref{fig:rate_MSE_Nk=00003DMk=00003D4_Me=00003D6}(a) and (b)
characterize the rate and the MSE performance, respectively, of our MT-MSE
algorithm in the presence of a sophisticated eavesdropper for the MIMO interference
channel, where we have $M=4$, $M_{e}=6$ and $\varepsilon_{\mathit{k}}=1.5$. The maximum DoF of 4
can indeed be achieved, which implies that each transmitter can attain a DoF of two\cite{Yetis_Feasibility_IA_TSP2010}.
In the MIMO interference network model considered, statistically there is no difference among the three legitimate transmitter-receiver pairs. Therefore, without any loss of generality we evaluate
both the secrecy rate and the MSE between Transmitter 1 and Receiver
1 as a representative of the legitimate transmitter-receiver pairs.
Fig. \ref{fig:rate_MSE_Nk=00003DMk=00003D4_Me=00003D6}(a) portrays
the achievable communication rate and secrecy rate versus the SNR for different
algorithms. As shown in Fig. \ref{fig:rate_MSE_Nk=00003DMk=00003D4_Me=00003D6}(a),
the MT-MSE algorithm is capable of supporting a useful positive secrecy rate, which increases
as the SNR increases. By contrast, the traditional IA algorithm \cite{Gomadam_TIT2011,Sasaki_Secure_IA_ISAP2012}
fails to guarantee secure communications in the scenario having a sophisticated
eavesdropper, since the secrecy rate achieved by the traditional
IA algorithm is zero. Although the achievable communication rate between the legitimate Transmitter 1 and Receiver 1
upon using the MT-MSE algorithm is lower than that of using the traditional IA algorithm\footnote{This is because our MT-MSE algorithm strikes a compromise between the attainable communication rate and secrecy rate.}, the MT-MSE algorithm
is capable of guaranteeing secure communications right across the entire SNR region. In other words,
the MT-MSE algorithm sacrifices a certain fraction of its potentially attainable communication rate for the sake of achieving a useful positive secrecy rate of data transmission.

Furthermore, we can see in Fig. \ref{fig:rate_MSE_Nk=00003DMk=00003D4_Me=00003D6}(b) that upon using our MT-MSE algorithm, the eavesdropper's MSE remains above the given threshold of $\varepsilon_{\mathit{k}}=1.5$, and it is higher than the MSE of the legitimate Receiver 1 under all the SNR conditions considered. We can also observe that the eavesdropper's MSE increases quite rapidly in the high SNR region, because the impact of the interference becomes more dominant than that of the noise. By contrast, both the MSE of the legitimate Receiver 1 and the total MSE of all the legitimate receivers decrease upon increasing the SNR. Hence, the communication rate of each legitimate transmitter-receiver pair becomes higher than that of the transmitter-eavesdropper link. As a beneficial result, secure communications can be realized.

\begin{figure*}[!t]
\begin{center}
\subfloat[Communication and secrecy rate for $M=4$ , $M_{e}=5$]{\includegraphics[width=3.5in]{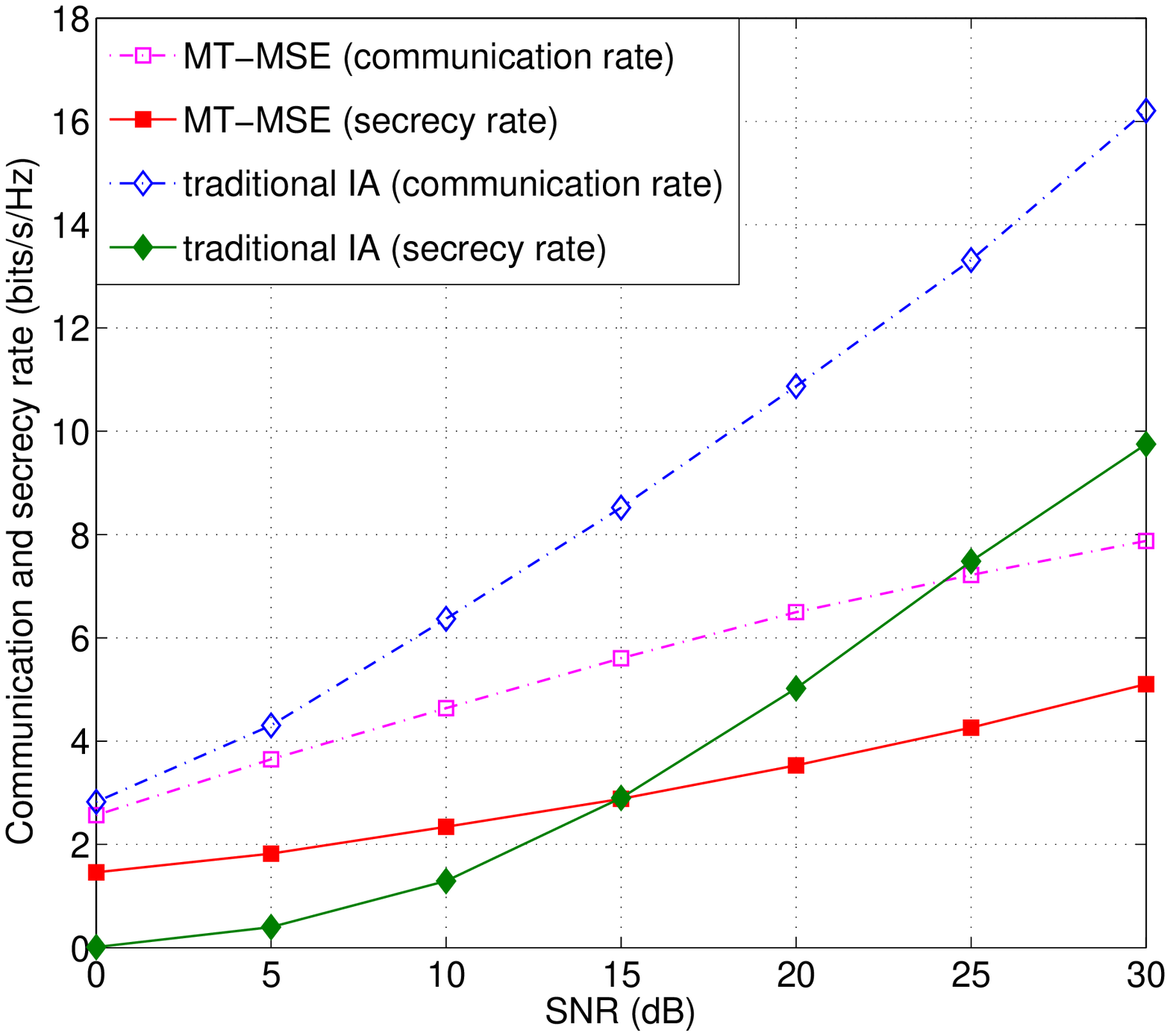}
\label{subfig:communication and secrecy rate Me=00003D5}}
\subfloat[Communication and secrecy rate for $M=4$, $M_{e}=4$]{\includegraphics[width=3.5in]{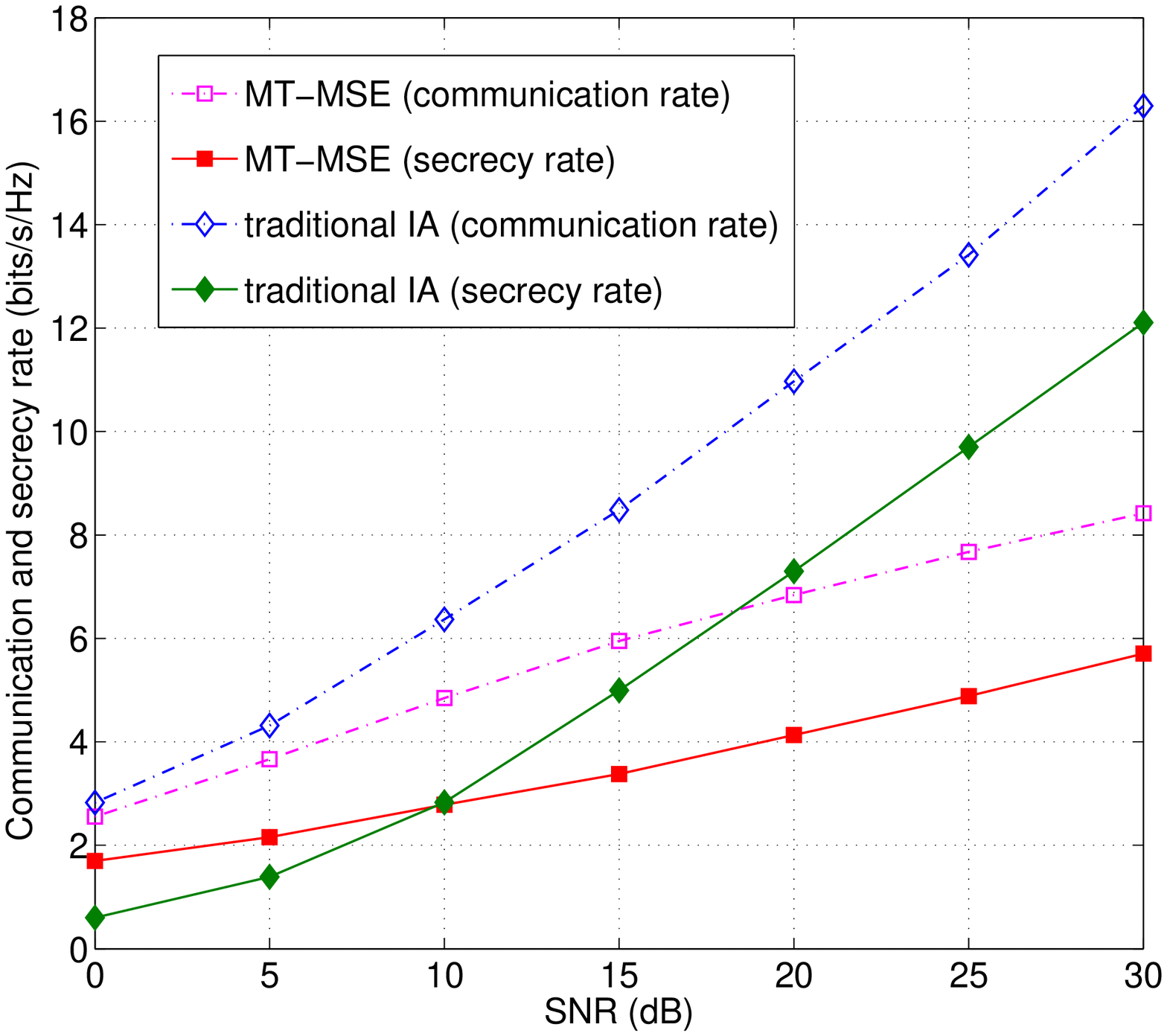}
\label{subfig:communication and secrecy rate Me=00003D4}}
\par
\end{center}
\raggedright{}\protect\caption{Communication rate and secrecy rate versus SNR performance comparison
of the proposed MT-MSE algorithm and the traditional IA algorithm \cite{Gomadam_TIT2011,Sasaki_Secure_IA_ISAP2012}
in the presence of an ``unsophisticated eavesdropper'', where we have $M=4$ and $\varepsilon_{\mathit{k}}=1.5$.}
\label{fig:rate__Nk=00003DMk=00003D4_Me=00003D4and5}
\end{figure*}

In addition to investigating the ``sophisticated eavesdropper'' scenario in Fig. \ref{fig:rate_MSE_Nk=00003DMk=00003D4_Me=00003D6}, in Fig. \ref{fig:rate__Nk=00003DMk=00003D4_Me=00003D4and5} we also quantify the performance of the MT-MSE algorithm in the presence of an ``unsophisticated eavesdropper'', which has an identical, lower or slightly higher number of antennas than each legitimate
transmitter and receiver. Similar observations can be made for this unsophisticated/weak eavesdropper scenario in the context of the MIMO interference channel, as shown in Fig. \ref{fig:rate__Nk=00003DMk=00003D4_Me=00003D4and5}(a)
and Fig. \ref{fig:rate__Nk=00003DMk=00003D4_Me=00003D4and5}(b), where the number of eavesdropper antennas is set to $M_{e}=4$ and $M_{e}=5$, respectively, while the MSE-threshold is set to $\varepsilon_{\mathit{k}}=1.5$. The results given in Fig. \ref{fig:rate__Nk=00003DMk=00003D4_Me=00003D4and5}(a)
and (b) indicate that the MT-MSE algorithm is capable of guaranteeing
secure communications for all the SNRs considered. In particular, although the traditional IA algorithm is secure in the scenario of having an unsophisticated eavesdropper, the proposed MT-MSE algorithm
exhibits a more appealing secrecy performance than the traditional IA algorithm in the low to medium SNR
regions. We observe from Fig. \ref{fig:rate__Nk=00003DMk=00003D4_Me=00003D4and5}(a)
and (b) that the SNR ``divide'' between the secrecy rates achieved by the MT-MSE algorithm and the traditional IA algorithm is about 15 dB and 10 dB, respectively. In fact, when
the number of antennas at each legitimate transmitter and receiver, i.e. $M$,
is no lower than that of the eavesdropper $M_e$, the traditional
IA algorithm remains secure, because the eavesdropper no longer has a sufficiently high number of spatial dimensions to facilitate successful signal detection in the IA system\footnote{In this scenario, excessive residual interference is imposed on the eavesdropper. This residual interference cannot be reduced by any arbitrary decoding matrices, since the number of dimensions required by decoding/detection is lower than the number of dimensions of the interference signal subspace. As a result, the IA system operating in the presence of an``unsophisticated eavesdropper'' is secure in nature. Therefore, in Fig. \ref{fig:rate__Nk=00003DMk=00003D4_Me=00003D4and5} we did not provide the numerical results for the scenario of $M > M_e$. }\cite{Sasaki_Secure_IA_ISAP2012}. This statement is true even when $M_e$ is slightly larger than $M$, for example, the configuration of ``$M_e = 5, M = 4$'' investigated in Fig. \ref{fig:rate__Nk=00003DMk=00003D4_Me=00003D4and5}(a).

\begin{figure*}[!t]
\begin{center}
\subfloat[Secrecy rate]{\includegraphics[width=3.5in]{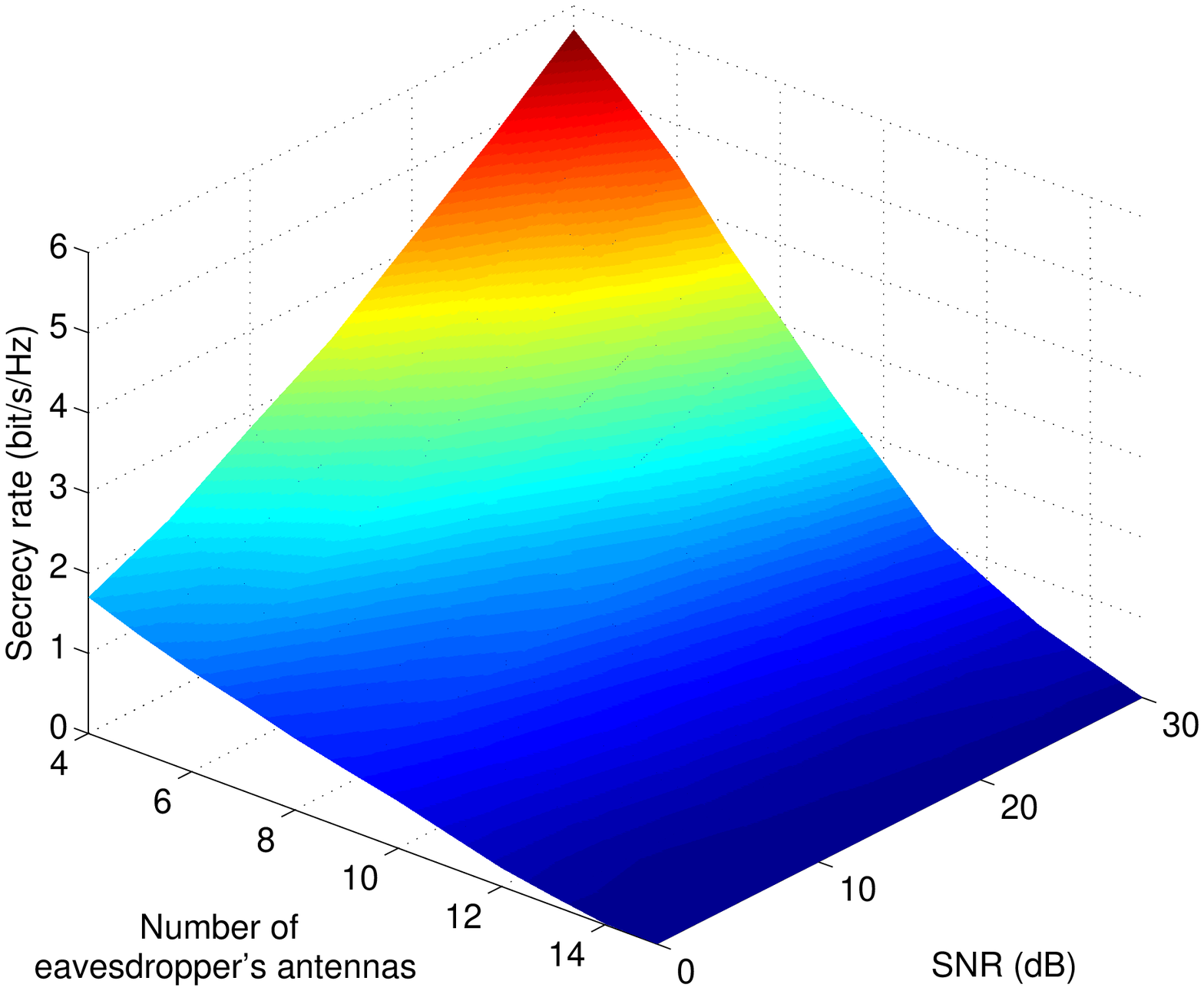}
\label{subfig:secrecy_rate_vs_eve_antennas}}
\subfloat[Maximum achievable received rate at the eavesdropper]{\includegraphics[width=3.5in]{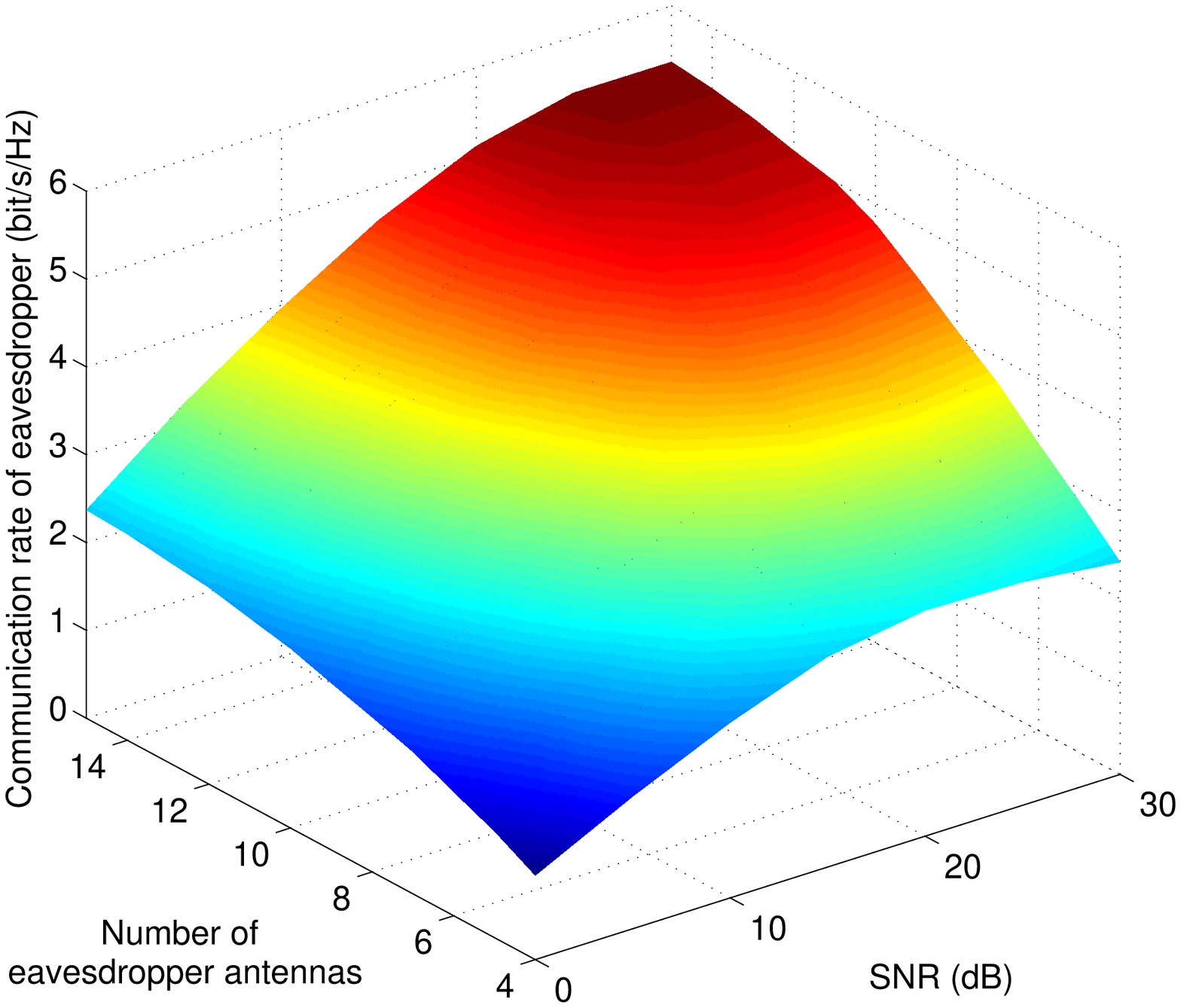}
\label{subfig:communication_rate_vs_eve_antennas}}
\par
\end{center}
\raggedright{}\protect\caption{The variation of the secrecy rate and the maximum achievable received rate between
a legitimate transmitter and the eavesdropper when the proposed MT-MSE algorithm is invoked, where we have $M = 4$ and $\varepsilon_{\mathit{k}}=1.5$. }
\label{fig:secrecy_rate_and_eve_rate}
\end{figure*}

Furthermore, in Fig. \ref{fig:secrecy_rate_and_eve_rate}(a) and (b) we
characterize the variation of the secrecy rate and the maximum achievable received rate between the legitimate Transmitter 1 and the eavesdropper, both as a function of the number of eavesdropper antennas $M_{e}$ and as that of the SNR, where the MT-MSE algorithm is employed and we have $M=4$ as well as $\varepsilon_{\mathit{k}}=1.5$. Fig. \ref{fig:secrecy_rate_and_eve_rate}(a)
portrays the variation of the secrecy rate. As shown in Fig. \ref{fig:secrecy_rate_and_eve_rate}(a),
the secrecy rate attained decreases as the number of eavesdropper antennas is increased and as the SNR is decreased. Furthermore, the decrease of the attainable secrecy rate becomes slow when the number of eavesdropper antennas is relatively large (e.g. $M_{e} \ge 8$), which implies that the achievable secrecy rate approaches the lower bound of zero as long as the number of eavesdropper antennas becomes sufficiently large. It can be observed that in general the MT-MSE algorithm remains secure even in the challenging scenario of having a sophisticated eavesdropper, which is equipped with a larger number of antennas (i.e. $ 4 < M_e <  15$) than each legitimate transmitter and receiver. However, even the MT-MSE algorithm fails to guarantee secure communications when the number of eavesdropper antennas satisfies $M_{e} \ge 15$. When the number of eavesdropper antennas is
increased, the number of dimensions available for signal detection/decoding at the eavesdropper is also
increased. In this situation, the eavesdropper becomes capable of simultaneously decoding multiple signals, which are mutually interfering with each other.  As a result, the eavesdropper becomes capable of wiretapping the transmitted
data and hence the achievable secrecy rate becomes zero.

Fig. \ref{fig:secrecy_rate_and_eve_rate}(b)
portrays the variation of the maximum achievable received rate between the legitimate
Transmitter 1 and the eavesdropper versus both the SNR and the number of eavesdropper antennas. It can be seen that this rate increases
as the number of eavesdropper antennas is increased, albeit the rate increases only
slowly when the number of eavesdropper antennas is relatively high. Additionally, this rate first increases substantially as the SNR increases, but after reaching its peak value the rate is slightly reduced at the high SNRs considered. This is because the eavesdropper MSE is increased when the SNR is high.

\begin{figure}
\begin{centering}
\includegraphics[width=3.5in]{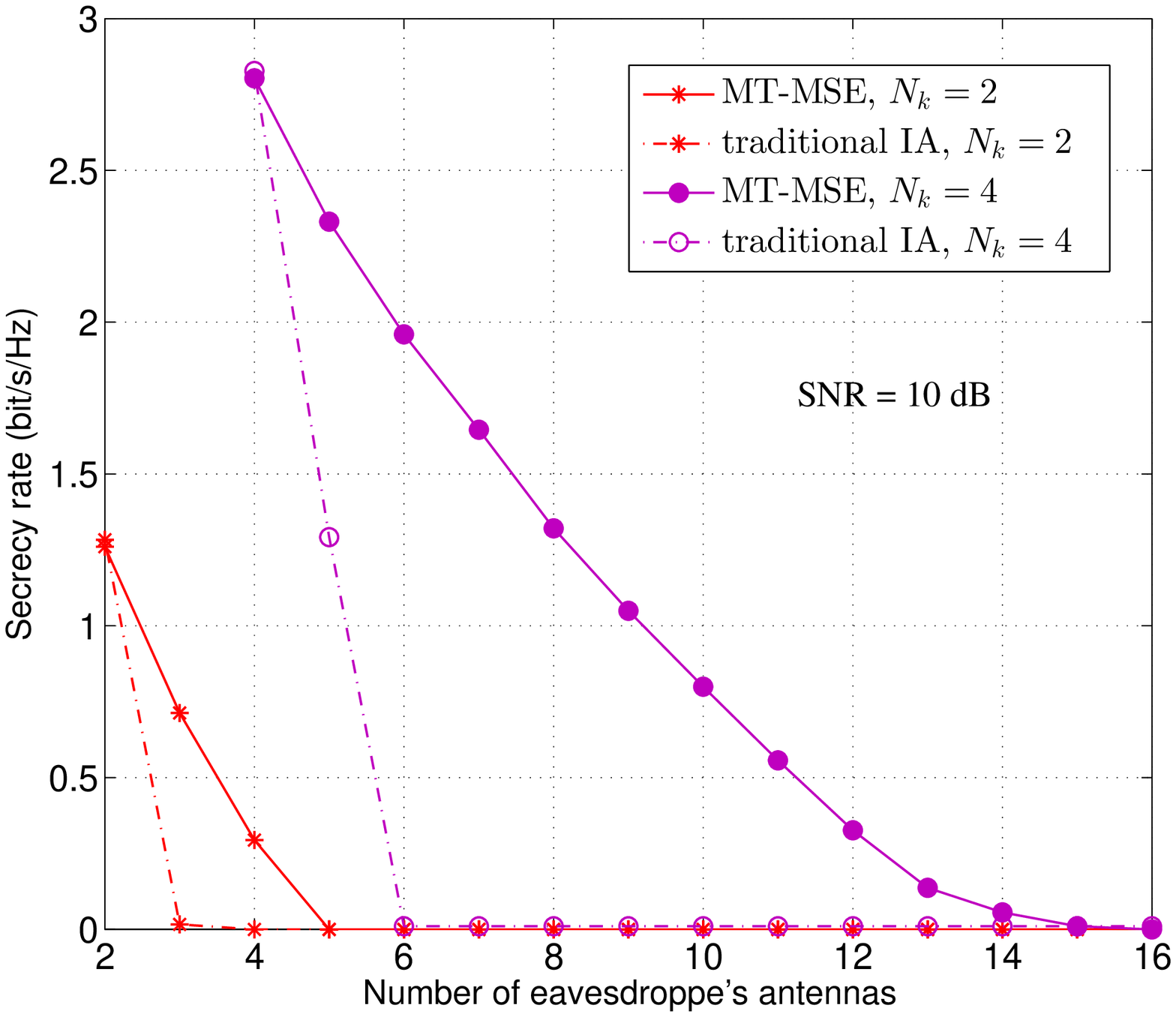}
\par\end{centering}
\protect\caption{Secrecy rate performance comparison versus the number of eavesdropper antennas for the case of SNR=10 dB.}
\raggedright{}\label{fig:secrecy_rate_10dB}
\end{figure}

Fig. \ref{fig:secrecy_rate_10dB} shows the secrecy rate comparison
between the MT-MSE algorithm and the traditional IA algorithm, when the number of eavesdropper antennas $M_{e}$ is varied. In Fig. \ref{fig:secrecy_rate_10dB}, the SNR is fixed to 10 dB and the number of antennas of each legitimate transmitter and
receiver is set to $M=2$ or $M=4$.
Again, as shown in Fig. \ref{fig:secrecy_rate_10dB}, the MT-MSE algorithm
outperforms the traditional IA algorithm, when the number of eavesdropper
antennas is higher than the number of antennas at each legitimate
transmitter and receiver. By contrast, they exhibit a similar performance when the number of eavesdropper antennas is equal to the number of antennas at each legitimate transmitter
and receiver. It can also be observed in Fig. \ref{fig:secrecy_rate_10dB} that the secrecy rate achieved by
the traditional IA algorithm decreases rapidly as the number of eavesdropper antennas $M_{e}$ is increased. However, the secrecy rate achieved by the MT-MSE algorithm is reduced much slower than that of the traditional IA
algorithm. This is because we recalculate both the TPC and receive filter matrices to guarantee that the eavesdropper's MSE is above the given threshold.  By contrast, the eavesdropper's MSE decreases upon increasing the SNR in the traditional IA approach. Numerically, for the $M=2$ or $M=4$ MIMO interference channel,
the secrecy rate achieved by the traditional IA algorithm reduces to
zero, when the number of eavesdropper antennas $M_{e}$ is higher
than 2 or 5, respectively. However, in the above MIMO interference channel, the
secrecy rate achieved by the MT-MSE algorithm is reduced to zero when
the number of eavesdropper antennas $M_{e}$ is higher than
4 or 15, respectively. In general, the MT-MSE algorithm has a more appealing security performance than the traditional
IA algorithm in the presence of a sophisticated eavesdropper.

\section{Conclusions}
\label{Section:Conclusion}
In this paper, we investigated secure communications for a $K$-user MIMO interference channel in the presence of an eavesdropper. An iterative distributed total MSE minimization algorithm, namely the MT-MSE algorithm of Section III, has been proposed for achieving secure communications. The proposed MT-MSE algorithm jointly designs the TPC and receive filter matrices by minimizing the total MSE of all receivers subject to specific secrecy and transmit power constraints. We also demonstrated the convergence of the proposed MT-MSE algorithm. Our simulation results demonstrated that secure communications can be guaranteed by using the proposed MT-MSE algorithm even in the challenging scenario of having a ``sophisticated/strong eavesdropper'', while the traditional IA algorithm becomes insecure in this  scenario. Our simulation results also showed that the proposed MT-MSE algorithm outperforms the traditional IA algorithm in both the low and medium SNR regions when an ``unsophisticated/weak eavesdropper'' was considered. In general, the proposed MT-MSE algorithm is capable of improving the secrecy rate performance of the MIMO interference channel compared to the traditional IA scheme.

\begin{appendices}
\section{Calculate $\hat{\lambda}_{k}^{\left(n+1\right)}$}
\label{Appendix A}
In Step 3-ii of MT-MSE algorithm, $\hat{\lambda}_{k}^{\left(n+1\right)}$
can be derived from the second constraint of (\ref{eq:MT_MSE_problem})
at the $\left(n+1\right)$-th iteration, i.e. solving the following
equation
\begin{equation}
\mathrm{Tr}\left[\mathbf{\hat{V}}_{k}^{\left(n+1\right)}\left(\mathbf{\hat{V}}_{k}^{\left(n+1\right)}\right)^{H}\right]=\mathbf{\mathrm{\mathit{p_{k}}}},\, k\in\left\{ 1,2,\ldots,K\right\} ,\label{eq:trace_VV^H}
\end{equation}
where $\hat{\mathbf{V}}_{k}^{\left(n+1\right)}$ is the function of the
variable $\hat{\lambda}_{k}^{\left(n+1\right)}$ and is given by
(\ref{eq:V^hat}). In (\ref{eq:V^hat}), $\hat{\mathbf{U}}_{l}^{\left(n+1\right)}$
and $\hat{\mathbf{U}}_{e,k}^{\left(n+1\right)}$ are updated at the
beginning of the $\left(n+1\right)$-th iteration, while $\mu_{k}^{\left(n\right)}$
is calculated during the $n$-th iteration. For the sake of convenience, we define:
\begin{align}
\mathbf{\boldsymbol{\Psi}_{\mathit{k}}^{\left(\mathit{n\mathrm{+1}}\right)}} & \triangleq\sum_{\mathit{k}=1}^{\mathit{K}}\left[\mathrm{\mathbf{H}_{\mathit{lk}}^{\mathit{H}}}\mathbf{U}_{l}^{\left(n+1\right)}\left(\mathbf{U}_{l}^{\left(n+1\right)}\right)^{H}\mathbf{H}_{\mathit{lk}}\right] \nonumber \\ & -\mu_{k}^{\left(n\right)}\mathrm{\mathbf{H}_{\mathit{ek}}^{\mathit{H}}}\mathbf{U}_{e,k}^{\left(n+1\right)}\left(\mathbf{U}_{e,k}^{\left(n+1\right)}\right)^{H}\mathbf{H}_{e\mathit{k}}\nonumber \\
 & \triangleq\mathbf{P}_{\mathit{k}}^{\left(\mathit{n\mathrm{+1}}\right)}\boldsymbol{\Sigma}_{\mathit{k}}^{\left(\mathit{n\mathrm{+1}}\right)}\left(\mathbf{P}_{\mathit{k}}^{\left(\mathit{n\mathrm{+1}}\right)}\right)^{H},\label{eq:Psi_k}
\end{align}
\begin{equation}
\mathbf{\boldsymbol{\Theta}}_{k}^{\left(n+1\right)}\triangleq\mathrm{\mathbf{H}_{\mathit{kk}}^{\mathit{H}}}\mathbf{U}_{k}^{\left(n+1\right)}-\mu_{k}^{\left(n\right)}\mathrm{\mathbf{H}_{\mathit{ek}}^{\mathit{H}}}\mathbf{U}_{e,k}^{\left(n+1\right)},\label{eq:Theta_k}
\end{equation}
where $\mathbf{P}_{\mathit{k}}^{\left(\mathit{n\mathrm{+1}}\right)}\boldsymbol{\Sigma}_{\mathit{k}}^{\left(\mathit{n\mathrm{+1}}\right)}\left(\mathbf{P}_{\mathit{k}}^{\left(\mathit{n\mathrm{+1}}\right)}\right)^{H}$
is the result of the singular value decomposition of the Hermitian matrix $\mathbf{\boldsymbol{\Psi}_{\mathit{k}}^{\left(\mathit{n\mathrm{+1}}\right)}}$.

Substituting (\ref{eq:V^hat}), (\ref{eq:Psi_k}) and (\ref{eq:Theta_k})
into the right-hand side of (\ref{eq:trace_VV^H}), we get \eqref{eq:lambda_k_hat},
\begin{figure*}[htbp]
\begin{multline}
\mathrm{Tr}\left[\mathbf{\hat{V}}_{k}^{\left(n+1\right)}\left(\mathbf{\hat{V}}_{k}^{\left(n+1\right)}\right)^{H}\right]\\
=\mathrm{Tr}\left\{ \left(\boldsymbol{\Psi}_{\mathit{k}}^{\left(\mathit{n\mathrm{+1}}\right)}+\hat{\lambda}_{k}^{\left(n+1\right)}\mathbf{I}\right)^{-1}\mathbf{\boldsymbol{\Theta}}_{k}^{\left(n+1\right)}\left[\left(\boldsymbol{\Psi}_{\mathit{k}}^{\left(\mathit{n\mathrm{+1}}\right)}+\hat{\lambda}_{k}^{\left(n+1\right)}\mathbf{I}\right)^{-1}\mathbf{\boldsymbol{\Theta}}_{k}^{\left(n+1\right)}\right]^{H}\right\} \qquad\qquad\qquad\qquad\qquad\qquad\qquad\qquad\quad\\
    =\mathrm{Tr}\left\{ \left(\mathbf{\boldsymbol{\Theta}}_{k}^{\left(n+1\right)}\right)^{H}\left[\mathbf{P}_{\mathit{k}}^{\left(\mathit{n\mathrm{+1}}\right)}\boldsymbol{\Sigma}_{\mathit{k}}^{\left(\mathit{n\mathrm{+1}}\right)}\left(\mathbf{P}_{\mathit{k}}^{\left(\mathit{n\mathrm{+1}}\right)}\right)^{H}+\hat{\lambda}_{k}^{\left(n+1\right)}\mathbf{I}\right]^{-1}\left[\mathbf{P}_{\mathit{k}}^{\left(\mathit{n\mathrm{+1}}\right)}\boldsymbol{\Sigma}_{\mathit{k}}^{\left(\mathit{n\mathrm{+1}}\right)}\left(\mathbf{P}_{\mathit{k}}^{\left(\mathit{n\mathrm{+1}}\right)}\right)^{H}+\hat{\lambda}_{k}^{\left(n+1\right)}\mathbf{I}\right]^{-1}\mathbf{\boldsymbol{\Theta}}_{k}^{\left(n+1\right)}\right\} \quad\\
=\left(\hat{\lambda}_{k}^{\left(n+1\right)}\right)^{-2}\mathrm{Tr}\left\{ \left(\mathbf{P}_{\mathit{k}}^{\left(\mathit{n\mathrm{+1}}\right)}\right)^{H}\mathbf{\boldsymbol{\Theta}}_{k}^{\left(n+1\right)}\left(\mathbf{\boldsymbol{\Theta}}_{k}^{\left(n+1\right)}\right)^{H}\mathbf{P}_{\mathit{k}}^{\left(\mathit{n\mathrm{+1}}\right)}\left\{ \mathbf{I}-\left[\hat{\lambda}_{k}^{\left(n+1\right)}\left(\boldsymbol{\Sigma}_{\mathit{k}}^{\left(\mathit{n\mathrm{+1}}\right)}\right)^{-1}+\mathbf{I}\right]^{-1}\right\} ^{2}\right\} \qquad\qquad\qquad\quad\\
=\sum_{i=1}^{N_{k}}\frac{\left[\left(\mathbf{P}_{\mathit{k}}^{\left(\mathit{n\mathrm{+1}}\right)}\right)^{H}\mathbf{\boldsymbol{\Theta}}_{k}^{\left(n+1\right)}\left(\mathbf{\boldsymbol{\Theta}}_{k}^{\left(n+1\right)}\right)^{H}\mathbf{P}_{\mathit{k}}^{\left(\mathit{n\mathrm{+1}}\right)}\right]_{\left[i,i\right]}}{\left[\hat{\lambda}_{k}^{\left(n+1\right)}+\left(\boldsymbol{\Sigma}_{\mathit{k}}^{\left(\mathit{n\mathrm{+1}}\right)}\right)_{\left[i,i\right]}\right]^{2}}=p_{k},\quad k\in\left\{ 1,2,\ldots,K\right\} .\qquad\qquad\qquad\qquad\qquad\qquad\qquad\label{eq:lambda_k_hat}
\end{multline}
\end{figure*}
where we have exploited the matrix inversion lemma\footnote{We have $\left(\mathbf{A}+\mathbf{CBC^{\mathit{T}}}\right)^{-1} =\mathbf{A^{\mathrm{-1}}} -\mathbf{A^{\mathrm{-1}}}\mathbf{C}\left(\mathbf{B^{\mathrm{-1}}}+\mathbf{C}^{\mathit{T}}\mathbf{A^{\mathrm{-1}}}\mathbf{C}\right)^{-1}\mathbf{C}^{\mathit{T}}\mathbf{A^{\mathrm{-1}}}$.} to simplify $\left[\mathbf{P}_{\mathit{k}}^{\left(\mathit{n\mathrm{+1}}\right)}\boldsymbol{\Sigma}_{\mathit{k}}^{\left(\mathit{n\mathrm{+1}}\right)}\left(\mathbf{P}_{\mathit{k}}^{\left(\mathit{n\mathrm{+1}}\right)}\right)^{H}+\hat{\lambda}_{k}^{\left(n+1\right)}\mathbf{I}\right]^{-1}$.
Therefore, it is found that $\hat{\lambda}_{k}^{\left(n+1\right)}$
can be easily solved according to (\ref{eq:lambda_k_hat}).

\section{Calculate $\widetilde{\mu}_{k}^{\left(n+1\right)}$}
\label{Appendix B}
In Step 6 of the MT-MSE algorithm,  after solving $\hat{\lambda}_{k}^{\left(n+1\right)}$,
we employ the updated $\lambda_{k}^{\left(n+1\right)}$ and intermediate
result $\mathbf{\bar{V}}_{k}^{\left(n+1\right)}$ given by (\ref{eq:intermediate_result_V})
to solve $\widetilde{\mu}_{k}^{\left(n+1\right)}$ from the first
constraint of (\ref{eq:MT_MSE_problem}) at the $\left(n+1\right)$-th
iteration. Here, we define $\mathbf{\boldsymbol{\Omega}_{\mathit{k}}^{\left(\mathit{n\mathrm{+1}}\right)}}$ as \eqref{eq:omega_k},
\begin{figure*}
\begin{align}
\mathbf{\boldsymbol{\Omega}_{\mathit{k}}^{\left(\mathit{n\mathrm{+1}}\right)}} & \triangleq\begin{cases}
\left(\mathbf{U}_{e,k}^{\left(n+1\right)}\right)^{H}\sum_{\mathit{i}=1}^{\mathit{k\mathrm{-1}}}\left\{ \mathrm{\mathbf{H}_{\mathit{ei}}}\mathbf{\bar{V}}_{\mathit{i}}^{\left(\mathit{n\mathrm{+1}}\right)}\left(\mathbf{\bar{V}}_{i}^{\left(\mathit{n\mathrm{+1}}\right)}\right)^{H}\mathrm{\mathbf{H}_{\mathit{ei}}^{\mathit{H}}}\right\} \mathbf{U}_{e,k}^{\left(n+1\right)}+\sigma_{e}^{2}\left(\mathbf{U}_{e,k}^{\left(n+1\right)}\right)^{H}\mathbf{U}_{e,k}^{\left(n+1\right)}+\mathbf{I},\: k=1, \\
\\
\left(\mathbf{U}_{e,k}^{\left(n+1\right)}\right)^{H}\sum_{\mathit{i}=1}^{\mathit{k\mathrm{-1}}}\left\{ \mathrm{\mathbf{H}_{\mathit{ei}}}\mathbf{V}_{\mathit{i}}^{\left(\mathit{n\mathrm{+1}}\right)}\left(\mathbf{V}_{i}^{\left(\mathit{n\mathrm{+1}}\right)}\right)^{H}\mathrm{\mathbf{H}_{\mathit{ei}}^{\mathit{H}}}\right\} \mathbf{U}_{e,k}^{\left(n+1\right)}+\sigma_{e}^{2}\left(\mathbf{U}_{e,k}^{\left(n+1\right)}\right)^{H}\mathbf{U}_{e,k}^{\left(n+1\right)}\\
+\left(\mathbf{U}_{e,k}^{\left(n+1\right)}\right)^{H}\sum_{\mathit{i}=k+1}^{\mathit{K}}\left\{ \mathrm{\mathbf{H}_{\mathit{ei}}}\mathbf{\bar{V}}_{\mathit{i}}^{\left(\mathit{n\mathrm{+1}}\right)}\left(\mathbf{\bar{V}}_{i}^{\left(\mathit{n\mathrm{+1}}\right)}\right)^{H}\mathrm{\mathbf{H}_{\mathit{ei}}^{\mathit{H}}}\right\} \mathbf{U}_{e,k}^{\left(n+1\right)}+\mathbf{I},\; k\in\left\{ 2,\ldots,K\right\}.
\end{cases}\label{eq:omega_k}
\end{align}
\end{figure*}
where $\mathbf{\bar{V}}_{i}^{\left(n+1\right)}$, $\mathbf{V}_{i}^{\left(n+1\right)}$,
$\mathbf{\boldsymbol{\Psi}_{\mathit{i}}^{\left(\mathit{n\mathrm{+1}}\right)}}$
and $\mathbf{\boldsymbol{\Theta}}_{i}^{\left(n+1\right)}$ are given
by (\ref{eq:intermediate_result_V}), (\ref{eq:updated_V}), (\ref{eq:Psi_k})
and (\ref{eq:Theta_k}), respectively. Let
\begin{align}
\mathbf{\boldsymbol{\Phi}}_{k}^{\left(n+1\right)} & \!\triangleq\!\sum_{l=1}^{K}\!\left[\mathrm{\mathbf{H}_{\mathit{lk}}^{\mathit{H}}}\mathbf{U}_{l}^{\left(n+1\right)}\left(\mathbf{U}_{l}^{\left(n+1\right)}\!\right)^{H}\mathrm{\mathbf{H}_{\mathit{lk}}}\right]\!+\!\lambda_{k}^{\left(n+1\!\right)}\mathbf{I},
\label{eq:Phi_k}
\end{align}
\begin{equation}
\mathbf{M}_{k}^{\left(n+1\right)}\triangleq\mathbf{H}_{\mathit{kk}}^{H}\mathbf{U}_{k}^{\left(n+1\right)},\label{eq:M_k}
\end{equation}
\begin{equation}
\mathbf{N}_{k}^{\left(n+1\right)}\triangleq\mathbf{H}_{\mathit{ek}}^{H}\mathbf{U}_{e,k}^{\left(n+1\right)}.\label{eq:N_k}
\end{equation}

Then, according to (\ref{eq:omega_k}) $\sim$ (\ref{eq:N_k}) and to the KKT
conditions of (\ref{eq:KKT_conditions-6}) $\sim$ (\ref{eq:KKT_conditions-8}),
the value of $\mathbf{\widetilde{\mu}}_{k}^{\left(n+1\right)}$ satisfying
the MSE constraint of (\ref{eq:MT_MSE_problem}) can be found as \eqref{eq:solving_mu_tilde_k_from_MSE_Eve},
\begin{figure*}
\begin{multline}
\mathbf{\mathrm{Tr}\left\{ \mathrm{\mathbf{E}_{Eve,TX_{\mathit{k}}}\left[\mathbf{\widetilde{V}}_{k}^{\left(\mathit{n}+1\right)}\left(\widetilde{\mu}_{k}^{\left(\mathit{n}+1\right)}\right)\right]}\right\} }\\
=\mathrm{Tr}\left[\mathbf{\boldsymbol{\Omega}_{\mathit{k}}^{\left(\mathit{n\mathrm{+1}}\right)}}+\left(\mathbf{N_{\mathit{k}}^{\left(\mathit{n\mathrm{+1}}\right)}}\right)^{H}\mathbf{\widetilde{V}}_{k}^{\left(\mathit{n}+1\right)}\left(\mathbf{\widetilde{V}}_{k}^{\left(\mathit{n}+1\right)}\right)^{H}\mathbf{N_{\mathit{k}}^{\left(\mathit{n\mathrm{+1}}\right)}}+\left(\mathbf{N_{\mathit{k}}^{\left(\mathit{n\mathrm{+1}}\right)}}\right)^{H}\mathbf{\widetilde{V}}_{k}^{\left(\mathit{n}+1\right)}+\left(\mathbf{\widetilde{V}}_{k}^{\left(\mathit{n}+1\right)}\right)^{H}\mathbf{N_{\mathit{k}}^{\left(\mathit{n\mathrm{+1}}\right)}}\right]=\varepsilon_{k}.\label{eq:solving_mu_tilde_k_from_MSE_Eve}
\end{multline}
\end{figure*}
where $\boldsymbol{\Omega}_{\mathit{k}}^{\left(\mathit{n\mathrm{+1}}\right)}$
is a predefined constant item with respect to $\mathbf{\widetilde{\mu}}_{k}^{\left(n+1\right)}$;
$\mathbf{\widetilde{V}}_{k}^{\left(\mathit{n}+1\right)}$ is the function
of $\mathbf{\widetilde{\mu}}_{k}^{\left(n+1\right)}$ and is given
by (\ref{eq:V^hat}). For the sake of convenience, we use $\mathbf{\widetilde{\mu}}_{k}$,
$\boldsymbol{\Omega}$, $\boldsymbol{\Phi}$, $\mathbf{M}$ and $\mathbf{N}$
instead of $\mathbf{\widetilde{\mu}}_{k}^{\left(n+1\right)}$, $\boldsymbol{\Omega}_{\mathit{k}}^{\left(\mathit{n\mathrm{+1}}\right)}$,
$\boldsymbol{\Phi}_{\mathit{k}}^{\left(\mathit{n\mathrm{+1}}\right)}$,
$\mathbf{M}_{k}^{\left(n+1\right)}$ and $\mathbf{N}_{k}^{\left(n+1\right)}$
in the following mathematical derivations, respectively. By employing
the matrix inversion lemma, $\mathbf{\widetilde{V}}_{k}^{\left(\mathit{n}+1\right)}$
in (\ref{eq:solving_mu_tilde_k_from_MSE_Eve}) can be rewritten as:
\begin{align}
\!\!\!\!\!\mathbf{\widetilde{V}}_{k}^{\!\left(\mathit{n}+1\!\right)}  
  \!\!\!=\! \!\!\left[\mathbf{\boldsymbol{\Phi}}^{\!\!-1}\!\!\!-\!\!\mathbf{\boldsymbol{\Phi}}^{\!\!-1}\!\mathbf{N}\mathbf{\boldsymbol{\Gamma}}\!\left(\!\mathbf{\boldsymbol{\Lambda}}\!\!-\!\!\mathbf{\widetilde{\mu}}_{k}^{-1}\!\mathbf{I}\right)^{-1}\!\!\mathbf{\boldsymbol{\Gamma}}^{H}\!\mathbf{N}^{H}\!\mathbf{\boldsymbol{\Phi}}^{-1}\!\right] \!\!\left(\mathbf{M}\!\!-\!\!\mathbf{\widetilde{\mu}}_{k}\mathbf{N}\right)\!,\label{eq:rewritten_V_tilde_k}
\end{align}
where $\boldsymbol{\Gamma}\mathbf{\boldsymbol{\Lambda}}\mathbf{\boldsymbol{\Gamma}}^{H}$
is the result of the singular value decomposition of the Hermitian matrix $\mathbf{N}^{H}\mathbf{\boldsymbol{\Phi}}^{-1}\mathbf{N}$.
For solving $\mathbf{\widetilde{\mu}}_{k}$ from (\ref{eq:solving_mu_tilde_k_from_MSE_Eve}),
it is readily observed that $\mathrm{Tr}\left[\boldsymbol{\Omega}\right]$,
$\mathrm{Tr}\left[\mathbf{N}^{H}\mathbf{\widetilde{V}}_{k}^{\left(\mathit{n}+1\right)}\left(\mathbf{\widetilde{V}}_{k}^{\left(\mathit{n}+1\right)}\right)^{H}\mathbf{N}\right]$,
$\mathrm{Tr}\left[\mathbf{N}^{H}\mathbf{\widetilde{V}}_{k}^{\left(\mathit{n}+1\right)}\right]$
and $\mathrm{Tr}\left[\left(\mathbf{\widetilde{V}}_{k}^{\left(\mathit{n}+1\right)}\right)^{H}\mathbf{N}\right]$
should be calculated. Furthermore, since $\mathrm{Tr}\left[\boldsymbol{\Omega}\right]$
is a constant item with respect to $\mathbf{\widetilde{\mu}}_{k}$,
we will elaborate on the other three items in the following.

Firstly, upon substituting (\ref{eq:rewritten_V_tilde_k}) into $\mathrm{Tr}\left[\mathbf{N}^{H}\mathbf{\widetilde{V}}_{k}^{\left(\mathit{n}+1\right)}\left(\mathbf{\widetilde{V}}_{k}^{\left(\mathit{n}+1\right)}\right)^{H}\mathbf{N}\right]$,
we arrive at \eqref{eq:first_item_solving_mu_k}.
\begin{figure*}
\begin{multline}
\mathrm{Tr}\left[\mathbf{N}^{H}\mathbf{\widetilde{V}}_{k}^{\left(\mathit{n}+1\right)}\left(\mathbf{\widetilde{V}}_{k}^{\left(\mathit{n}+1\right)}\right)^{H}\mathbf{N}\right]\\
=\mathrm{Tr}\left\{ \mathbf{N}^{H}\left[\mathbf{\boldsymbol{\Phi}}^{-1}-\mathbf{\boldsymbol{\Phi}}^{-1}\mathbf{N}\mathbf{\boldsymbol{\Gamma}}\left(\mathbf{\boldsymbol{\Lambda}}-\mathbf{\widetilde{\mu}}_{k}^{-1}\mathbf{I}\right)^{-1}\mathbf{\boldsymbol{\Gamma}}^{H}\mathbf{N}^{H}\mathbf{\boldsymbol{\Phi}}^{-1}\right]\left(\mathbf{M}-\mathbf{\widetilde{\mu}}_{k}\mathbf{N}\right)\left(\mathbf{M}^{H}-\mathbf{\widetilde{\mu}}_{k}\mathbf{N}^{H}\right)\left[\mathbf{\boldsymbol{\Phi}}^{-1}\qquad\qquad\qquad\quad\,\right.\right.\\
\left.\left.-\mathbf{\boldsymbol{\Phi}}^{-1}\mathbf{N}\mathbf{\boldsymbol{\Gamma}}\left(\mathbf{\boldsymbol{\Lambda}}-\mathbf{\widetilde{\mu}}_{k}^{-1}\mathbf{I}\right)^{-1}\mathbf{\boldsymbol{\Gamma}}^{H}\mathbf{N}^{H}\mathbf{\boldsymbol{\Phi}}^{-1}\right]\right\} \\
=\mathrm{Tr}\left\{ \mathbf{M}^{H}\left[\mathbf{\boldsymbol{\Phi}}^{-1}\mathbf{N}-\mathbf{\boldsymbol{\Phi}}^{-1}\mathbf{N}\mathbf{\boldsymbol{\Gamma}}\left(\mathbf{\boldsymbol{\Lambda}}-\mathbf{\widetilde{\mu}}_{k}^{-1}\mathbf{I}\right)^{-1}\mathbf{\boldsymbol{\Lambda}}\mathbf{\boldsymbol{\Gamma}}^{H}\right]\left[\mathbf{N}^{H}\mathbf{\boldsymbol{\Phi}}^{-1}-\mathbf{\boldsymbol{\Gamma}}\mathbf{\boldsymbol{\Lambda}}\left(\mathbf{\boldsymbol{\Lambda}}-\mathbf{\widetilde{\mu}}_{k}^{-1}\mathbf{I}\right)^{-1}\mathbf{\boldsymbol{\Gamma}}^{H}\mathbf{N}^{H}\mathbf{\boldsymbol{\Phi}}^{-1}\right]\mathbf{M}\right\} \qquad\quad\;\\
-\mathbf{\widetilde{\mu}}_{k}\mathrm{Tr}\left\{ \mathbf{M}^{H}\left[\mathbf{\boldsymbol{\Phi}}^{-1}\mathbf{N}-\mathbf{\boldsymbol{\Phi}}^{-1}\mathbf{N}\mathbf{\boldsymbol{\Gamma}}\left(\mathbf{\boldsymbol{\Lambda}}-\mathbf{\widetilde{\mu}}_{k}^{-1}\mathbf{I}\right)^{-1}\mathbf{\boldsymbol{\Lambda}}\mathbf{\boldsymbol{\Gamma}}^{H}\right]\left[\boldsymbol{\Gamma}\mathbf{\boldsymbol{\Lambda}}\mathbf{\boldsymbol{\Gamma}}^{H}-\mathbf{\boldsymbol{\Gamma}}\mathbf{\boldsymbol{\Lambda}}\left(\mathbf{\boldsymbol{\Lambda}}-\mathbf{\widetilde{\mu}}_{k}^{-1}\mathbf{I}\right)^{-1}\mathbf{\boldsymbol{\Lambda}}\mathbf{\boldsymbol{\Gamma}}^{H}\right]\right\} \\
-\mathbf{\widetilde{\mu}}_{k}\mathrm{Tr}\left\{ \left[\boldsymbol{\Gamma}\mathbf{\boldsymbol{\Lambda}}\mathbf{\boldsymbol{\Gamma}}^{H}-\mathbf{\boldsymbol{\Gamma}}\mathbf{\boldsymbol{\Lambda}}\left(\mathbf{\boldsymbol{\Lambda}}-\mathbf{\widetilde{\mu}}_{k}^{-1}\mathbf{I}\right)^{-1}\mathbf{\boldsymbol{\Lambda}}\mathbf{\boldsymbol{\Gamma}}^{H}\right]\left[\mathbf{N}^{H}\mathbf{\boldsymbol{\Phi}}^{-1}-\mathbf{\boldsymbol{\Gamma}}\mathbf{\boldsymbol{\Lambda}}\left(\mathbf{\boldsymbol{\Lambda}}-\mathbf{\widetilde{\mu}}_{k}^{-1}\mathbf{I}\right)^{-1}\mathbf{\boldsymbol{\Gamma}}^{H}\mathbf{N}^{H}\mathbf{\boldsymbol{\Phi}}^{-1}\right]\mathbf{M}\right\} \\
+\mathbf{\widetilde{\mu}}_{k}^{2}\mathrm{Tr}\left\{ \left[\boldsymbol{\Gamma}\mathbf{\boldsymbol{\Lambda}}\mathbf{\boldsymbol{\Gamma}}^{H}-\mathbf{\boldsymbol{\Gamma}}\mathbf{\boldsymbol{\Lambda}}\left(\mathbf{\boldsymbol{\Lambda}}-\mathbf{\widetilde{\mu}}_{k}^{-1}\mathbf{I}\right)^{-1}\mathbf{\boldsymbol{\Lambda}}\mathbf{\boldsymbol{\Gamma}}^{H}\right]\left[\boldsymbol{\Gamma}\mathbf{\boldsymbol{\Lambda}}\mathbf{\boldsymbol{\Gamma}}^{H}-\mathbf{\boldsymbol{\Gamma}}\mathbf{\boldsymbol{\Lambda}}\left(\mathbf{\boldsymbol{\Lambda}}-\mathbf{\widetilde{\mu}}_{k}^{-1}\mathbf{I}\right)^{-1}\mathbf{\boldsymbol{\Lambda}}\mathbf{\boldsymbol{\Gamma}}^{H}\right]\right\} .\label{eq:first_item_solving_mu_k}
\end{multline}
\end{figure*}
\!We \!further \!simplify \!the \!first \!item \!of \!(\ref{eq:first_item_solving_mu_k}) \!as \eqref{eq:first_item_of_firt_item_solving_mu_k},
\begin{figure*}
\begin{multline}
\mathrm{Tr}\left\{ \mathbf{M}^{H}\left[\mathbf{\boldsymbol{\Phi}}^{-1}\mathbf{N}-\mathbf{\boldsymbol{\Phi}}^{-1}\mathbf{N}\mathbf{\boldsymbol{\Gamma}}\left(\mathbf{\boldsymbol{\Lambda}}-\mathbf{\widetilde{\mu}}_{k}^{-1}\mathbf{I}\right)^{-1}\mathbf{\boldsymbol{\Lambda}}\mathbf{\boldsymbol{\Gamma}}^{H}\right]\left[\mathbf{N}^{H}\mathbf{\boldsymbol{\Phi}}^{-1}-\mathbf{\boldsymbol{\Gamma}}\mathbf{\boldsymbol{\Lambda}}\left(\mathbf{\boldsymbol{\Lambda}}-\mathbf{\widetilde{\mu}}_{k}^{-1}\mathbf{I}\right)^{-1}\mathbf{\boldsymbol{\Gamma}}^{H}\mathbf{N}^{H}\mathbf{\boldsymbol{\Phi}}^{-1}\right]\mathbf{M}\right\} \\
\!\!=\mathrm{Tr}\left(\boldsymbol{\Xi}\right)-2\mathrm{Tr}\left[\mathbf{\boldsymbol{\Lambda}}\left(\mathbf{\boldsymbol{\Lambda}}-\mathbf{\widetilde{\mu}}_{k}^{-1}\mathbf{I}\right)^{-1}\mathbf{\boldsymbol{\Gamma}}^{H}\boldsymbol{\Xi}\mathbf{\boldsymbol{\Gamma}}\right]+\mathrm{Tr}\left[\mathbf{\boldsymbol{\Lambda}^{\mathrm{2}}}\left(\mathbf{\boldsymbol{\Lambda}}-\mathbf{\widetilde{\mu}}_{k}^{-1}\mathbf{I}\right)^{-2}\mathbf{\boldsymbol{\Gamma}}^{H}\boldsymbol{\Xi}\mathbf{\boldsymbol{\Gamma}}\right]\qquad\qquad\qquad\qquad\qquad\qquad\qquad\:\:\:\\
\!\!=\sum_{i=1}^{d_{k}}\boldsymbol{\Xi}_{\left[i,i\right]}-\sum_{i=1}^{d_{k}}\left(\mathbf{\boldsymbol{\Gamma}}^{H}\boldsymbol{\Xi}\boldsymbol{\Gamma}\right)_{\left[i,i\right]}+\sum_{i=1}^{d_{k}}\left(\mathbf{\boldsymbol{\Gamma}}^{H}\boldsymbol{\Xi}\boldsymbol{\Gamma}\right)_{\left[i,i\right]}\left(\frac{1}{\mathbf{\widetilde{\mu}}_{k}\mathbf{\boldsymbol{\Lambda}_{\left[\mathrm{\mathit{i},\mathit{i}}\right]}}-1}\right)^{2}.\qquad\qquad\qquad\qquad\qquad\qquad\qquad\qquad\quad\;\label{eq:first_item_of_firt_item_solving_mu_k}
\end{multline}
\end{figure*}
where $\boldsymbol{\Xi}\triangleq\mathbf{\mathbf{N}}^{H}\mathbf{\boldsymbol{\Phi}}^{-1}\mathbf{M}^{H}\mathbf{M}\mathbf{\boldsymbol{\Phi}}^{-1}\mathbf{N}$.
Furthermore, the second item of (\ref{eq:first_item_solving_mu_k}) can be
simplified as \eqref{eq:second_item_of_firt_item_solving_mu_k},
\begin{figure*}
\begin{multline}
-\mathbf{\widetilde{\mu}}_{k}\mathrm{Tr}\left\{ \mathbf{M}^{H}\left[\mathbf{\boldsymbol{\Phi}}^{-1}\mathbf{N}-\mathbf{\boldsymbol{\Phi}}^{-1}\mathbf{N}\mathbf{\boldsymbol{\Gamma}}\left(\mathbf{\boldsymbol{\Lambda}}-\mathbf{\widetilde{\mu}}_{k}^{-1}\mathbf{I}\right)^{-1}\mathbf{\boldsymbol{\Lambda}}\mathbf{\boldsymbol{\Gamma}}^{H}\right]\left[\mathbf{\boldsymbol{\Gamma}}^{H}\mathbf{\boldsymbol{\Lambda}}\mathbf{\boldsymbol{\Gamma}}-\mathbf{\boldsymbol{\Gamma}}\mathbf{\boldsymbol{\Lambda}}\left(\mathbf{\boldsymbol{\Lambda}}-\mathbf{\widetilde{\mu}}_{k}^{-1}\mathbf{I}\right)^{-1}\mathbf{\boldsymbol{\Lambda}}\mathbf{\boldsymbol{\Gamma}}^{H}\right]\right\} \\
=-\mathbf{\widetilde{\mu}}_{k}\mathrm{Tr}\left[\mathbf{\boldsymbol{\Lambda}}\mathbf{\boldsymbol{\Gamma}}^{H}\mathbf{M}^{H}\mathbf{\boldsymbol{\Phi}}^{-1}\mathbf{N}\boldsymbol{\Gamma}-\mathbf{\boldsymbol{\Lambda}}\mathbf{\boldsymbol{\Gamma}}^{H}\mathbf{M}^{H}\mathbf{\boldsymbol{\Phi}}^{-1}\mathbf{N}\boldsymbol{\Gamma}\mathbf{\boldsymbol{\Lambda}}\left(\mathbf{\boldsymbol{\Lambda}}-\mathbf{\widetilde{\mu}}_{k}^{-1}\mathbf{I}\right)^{-1}-\mathbf{\boldsymbol{\Lambda}^{\mathrm{2}}}\mathbf{\boldsymbol{\Gamma}}^{H}\mathbf{M}^{H}\mathbf{\boldsymbol{\Phi}}^{-1}\mathbf{N}\boldsymbol{\Gamma}\left(\mathbf{\boldsymbol{\Lambda}}-\mathbf{\widetilde{\mu}}_{k}^{-1}\mathbf{I}\right)^{-1}\right.\qquad\;\\
\left.+\mathbf{\boldsymbol{\Lambda}}\mathbf{\boldsymbol{\Gamma}}^{H}\mathbf{M}^{H}\mathbf{\boldsymbol{\Phi}}^{-1}\mathbf{N}\boldsymbol{\Gamma}\left(\mathbf{\boldsymbol{\Lambda}}-\mathbf{\widetilde{\mu}}_{k}^{-1}\mathbf{I}\right)^{-1}\mathbf{\boldsymbol{\Lambda}^{\mathrm{2}}}\left(\mathbf{\boldsymbol{\Lambda}}-\mathbf{\widetilde{\mu}}_{k}^{-1}\mathbf{I}\right)^{-1}\right]\\
=-\mathbf{\widetilde{\mu}}_{k}\left\{ \mathrm{Tr}\left(\mathbf{\boldsymbol{\Lambda}}\boldsymbol{\Pi}\right)-\mathrm{Tr}\left[\left(\mathbf{\boldsymbol{\Lambda}}-\mathbf{\widetilde{\mu}}_{k}^{-1}\mathbf{I}\right)^{-1}\mathbf{\boldsymbol{\Lambda}^{\mathrm{2}}}\boldsymbol{\Pi}\right]-\mathrm{Tr}\left[\left(\mathbf{\boldsymbol{\Lambda}}-\mathbf{\widetilde{\mu}}_{k}^{-1}\mathbf{I}\right)^{-1}\mathbf{\boldsymbol{\Lambda}^{\mathrm{2}}}\boldsymbol{\Pi}\right]+\mathrm{Tr}\left[\mathbf{\boldsymbol{\Lambda}^{\mathrm{2}}}\left(\mathbf{\boldsymbol{\Lambda}}-\mathbf{\widetilde{\mu}}_{k}^{-1}\mathbf{I}\right)^{-2}\mathbf{\boldsymbol{\Lambda}}\boldsymbol{\Pi}\right]\right\} \quad\:\\
=-\mathbf{\widetilde{\mu}}_{k}\sum_{i=1}^{d_{k}}\left(\mathbf{\boldsymbol{\Lambda}}\boldsymbol{\Pi}\right)_{\left[i,i\right]}\left(\frac{1}{\mathbf{\widetilde{\mu}}_{k}\mathbf{\boldsymbol{\Lambda}_{\left[\mathrm{\mathit{i},\mathit{i}}\right]}}-1}\right)^{2}. \qquad\qquad\qquad\qquad\qquad\qquad\qquad\qquad\qquad\qquad\qquad\qquad\qquad\qquad\quad\;\;\label{eq:second_item_of_firt_item_solving_mu_k}
\end{multline}
\end{figure*}
where $\boldsymbol{\Pi}\triangleq\mathbf{\boldsymbol{\Gamma}}^{H}\mathbf{M}^{H}\mathbf{\boldsymbol{\Phi}}^{-1}\mathbf{N}\boldsymbol{\Gamma}$.
Similarly, since the third item of (\ref{eq:first_item_solving_mu_k})
is the conjugate transpose of (\ref{eq:second_item_of_firt_item_solving_mu_k}), we get \eqref{eq:third_item_of_firt_item_solving_mu_k}.
\begin{figure*}
\begin{multline}
-\mathbf{\widetilde{\mu}}_{k}\mathrm{Tr}\left\{ \left[\mathbf{\boldsymbol{\Gamma}}^{H}\mathbf{\boldsymbol{\Lambda}}\mathbf{\boldsymbol{\Gamma}}-\mathbf{\boldsymbol{\Gamma}}\mathbf{\boldsymbol{\Lambda}}\left(\mathbf{\boldsymbol{\Lambda}}-\mathbf{\widetilde{\mu}}_{k}^{-1}\mathbf{I}\right)^{-1}\mathbf{\boldsymbol{\Lambda}}\mathbf{\boldsymbol{\Gamma}}^{H}\right]\left[\mathbf{N}^{H}\mathbf{\boldsymbol{\Phi}}^{-1}-\mathbf{\boldsymbol{\Gamma}}^{H}\mathbf{\boldsymbol{\Lambda}}\left(\mathbf{\boldsymbol{\Lambda}}-\mathbf{\widetilde{\mu}}_{k}^{-1}\mathbf{I}\right)^{-1}\mathbf{\boldsymbol{\Gamma}}^{H}\mathbf{N}^{H}\mathbf{\boldsymbol{\Phi}}^{-1}\right]\mathbf{M}\right\} \\
=-\mathbf{\widetilde{\mu}}_{k}\sum_{i=1}^{d_{k}}\left(\boldsymbol{\Pi}^{H}\mathbf{\boldsymbol{\Lambda}}\right)_{\left[i,i\right]}\left(\frac{1}{\mathbf{\widetilde{\mu}}_{k}\mathbf{\boldsymbol{\Lambda}_{\left[\mathrm{\mathit{i},\mathit{i}}\right]}}-1}\right)^{2}.\qquad\qquad\qquad\qquad\qquad\qquad\qquad\qquad\qquad\qquad\qquad\qquad\qquad\quad\:\:\:\:\label{eq:third_item_of_firt_item_solving_mu_k}
\end{multline}
\end{figure*}
The last item of (\ref{eq:first_item_solving_mu_k}) is derived as \eqref{eq:fourth_item_of_firt_item_solving_mu_k}.
\begin{figure*}
\begin{multline}
\mathbf{\widetilde{\mu}}_{k}^{2}\mathrm{Tr}\left\{ \left[\mathbf{\boldsymbol{\Gamma}}^{H}\mathbf{\boldsymbol{\Lambda}}\mathbf{\boldsymbol{\Gamma}}-\mathbf{\boldsymbol{\Gamma}}\mathbf{\boldsymbol{\Lambda}}\left(\mathbf{\boldsymbol{\Lambda}}-\mathbf{\widetilde{\mu}}_{k}^{-1}\mathbf{I}\right)^{-1}\mathbf{\boldsymbol{\Lambda}}\mathbf{\boldsymbol{\Gamma}}^{H}\right]\left[\mathbf{\boldsymbol{\Gamma}}^{H}\mathbf{\boldsymbol{\Lambda}}\mathbf{\boldsymbol{\Gamma}}-\mathbf{\boldsymbol{\Gamma}}\mathbf{\boldsymbol{\Lambda}}\left(\mathbf{\boldsymbol{\Lambda}}-\mathbf{\widetilde{\mu}}_{k}^{-1}\mathbf{I}\right)^{-1}\mathbf{\boldsymbol{\Lambda}}\mathbf{\boldsymbol{\Gamma}}^{H}\right]\right\}\\
=\mathbf{\widetilde{\mu}}_{k}^{2}\left\{ \mathrm{Tr}\left(\mathbf{\boldsymbol{\Lambda}^{\mathrm{2}}}\right)-2\mathrm{Tr}\left[\mathbf{\boldsymbol{\Lambda}^{\mathrm{3}}}\left(\mathbf{\boldsymbol{\Lambda}}-\mathbf{\widetilde{\mu}}_{k}^{-1}\mathbf{I}\right)^{-1}\right]-\mathrm{Tr}\left[\mathbf{\boldsymbol{\Lambda}^{\mathrm{4}}}\left(\mathbf{\boldsymbol{\Lambda}}-\mathbf{\widetilde{\mu}}_{k}^{-1}\mathbf{I}\right)^{-2}\right]\right\} 
=\mathbf{\widetilde{\mu}}_{k}^{2}\sum_{i=1}^{d_{k}}\left(\mathbf{\boldsymbol{\Lambda}^{\mathrm{2}}}\right)_{\left[i,i\right]}\left(\frac{1}{\mathbf{\widetilde{\mu}}_{k}\mathbf{\boldsymbol{\Lambda}_{\left[\mathrm{\mathit{i},\mathit{i}}\right]}}-1}\right)^{2}.\label{eq:fourth_item_of_firt_item_solving_mu_k}
\end{multline}
\end{figure*}
By substituting (\ref{eq:first_item_of_firt_item_solving_mu_k})-(\ref{eq:fourth_item_of_firt_item_solving_mu_k}) into (\ref{eq:first_item_solving_mu_k}),
we arrive at \eqref{eq:result_first_item_solving_mu_k}.
\begin{figure*}
\begin{multline}
\mathrm{Tr}\left[\mathbf{N}^{H}\mathbf{\widetilde{V}}_{k}^{\left(\mathit{n}+1\right)}\left(\mathbf{\widetilde{V}}_{k}^{\left(\mathit{n}+1\right)}\right)^{H}\mathbf{N}\right]\\
=\sum_{i=1}^{d_{k}}\boldsymbol{\Xi}_{\left[i,i\right]}-\sum_{i=1}^{d_{k}}\left(\mathbf{\boldsymbol{\Gamma}}^{H}\boldsymbol{\Xi}\boldsymbol{\Gamma}\right)_{\left[i,i\right]}+\sum_{i=1}^{d_{k}}\left(\mathbf{\boldsymbol{\Gamma}}^{H}\boldsymbol{\Xi}\boldsymbol{\Gamma}\right)_{\left[i,i\right]}\left(\frac{1}{\mathbf{\widetilde{\mu}}_{k}\mathbf{\boldsymbol{\Lambda}_{\left[\mathrm{\mathit{i},\mathit{i}}\right]}}-1}\right)^{2}-\mathbf{\widetilde{\mu}}_{k}\sum_{i=1}^{d_{k}}\left(\mathbf{\boldsymbol{\Lambda}}\boldsymbol{\Pi}\right)_{\left[i,i\right]}\left(\frac{1}{\mathbf{\widetilde{\mu}}_{k}\mathbf{\boldsymbol{\Lambda}_{\left[\mathrm{\mathit{i},\mathit{i}}\right]}}-1}\right)^{2}\qquad\;\\
-\mathbf{\widetilde{\mu}}_{k}\sum_{i=1}^{d_{k}}\left(\boldsymbol{\Pi}^{H}\mathbf{\boldsymbol{\Lambda}}\right)_{\left[i,i\right]}\left(\frac{1}{\mathbf{\widetilde{\mu}}_{k}\mathbf{\boldsymbol{\Lambda}_{\left[\mathrm{\mathit{i},\mathit{i}}\right]}}-1}\right)^{2}+\mathbf{\widetilde{\mu}}_{k}^{2}\sum_{i=1}^{d_{k}}\left(\mathbf{\boldsymbol{\Lambda}^{\mathrm{2}}}\right)_{\left[i,i\right]}\left(\frac{1}{\mathbf{\widetilde{\mu}}_{k}\mathbf{\boldsymbol{\Lambda}_{\left[\mathrm{\mathit{i},\mathit{i}}\right]}}-1}\right)^{2}.\label{eq:result_first_item_solving_mu_k}
\end{multline}
\end{figure*}

Secondly, upon substituting (\ref{eq:rewritten_V_tilde_k}) into $\mathrm{Tr}\left[\mathbf{N}^{H}\mathbf{\widetilde{V}}_{k}^{\left(\mathit{n}+1\right)}\right]$, we have \eqref{eq:result_second_item_solving_mu_k},
\begin{figure*}
\begin{multline}
\mathrm{Tr}\left(\mathbf{N}^{H}\mathbf{\widetilde{V}}_{k}^{\left(\mathit{n}+1\right)}\right)
=\mathrm{Tr}\left\{ \mathbf{N}^{H}\left[\mathbf{\boldsymbol{\Phi}}^{-1}-\mathbf{\boldsymbol{\Phi}}^{-1}\mathbf{N}\mathbf{\boldsymbol{\Gamma}}\left(\mathbf{\boldsymbol{\Lambda}}-\mathbf{\widetilde{\mu}}_{k}^{-1}\mathbf{I}\right)^{-1}\mathbf{\boldsymbol{\Gamma}}^{H}\mathbf{N}^{H}\mathbf{\boldsymbol{\Phi}}^{-1}\right]\left(\mathbf{M}-\mathbf{\widetilde{\mu}}_{k}\mathbf{N}\right)\right\} \qquad\qquad\qquad\qquad\qquad\qquad\qquad\qquad\\
= \mathrm{Tr}\left(\boldsymbol{\Delta}\right)-\mathrm{Tr}\left[\boldsymbol{\Lambda}\left(\mathbf{\boldsymbol{\Lambda}}-\mathbf{\widetilde{\mu}}_{k}^{-1}\mathbf{I}\right)^{-1}\mathbf{\boldsymbol{\Gamma}}^{H}\boldsymbol{\Delta}\mathbf{\boldsymbol{\Gamma}}\right]-\mathbf{\widetilde{\mu}}_{k}\mathrm{Tr}\left(\boldsymbol{\Lambda}\right)+\mathbf{\widetilde{\mu}}_{k}\mathrm{Tr}\left[\mathbf{\boldsymbol{\Lambda}^{\mathrm{2}}}\left(\mathbf{\boldsymbol{\Lambda}}-\mathbf{\widetilde{\mu}}_{k}^{-1}\mathbf{I}\right)^{-1}\right]
\\=\sum_{i=1}^{d_{k}}\boldsymbol{\Delta}_{\left[i,i\right]}+\sum_{i=1}^{d_{k}}\mathbf{\boldsymbol{\Lambda}_{\left[\mathrm{\mathit{i},\mathit{i}}\right]}}\frac{1-\left(\mathbf{\boldsymbol{\Gamma}}^{H}\boldsymbol{\Delta}\mathbf{\boldsymbol{\Gamma}}\right)_{\left[\mathrm{\mathit{i},\mathit{i}}\right]}}{\mathbf{\widetilde{\mu}}_{k}\mathbf{\boldsymbol{\Lambda}_{\left[\mathrm{\mathit{i},\mathit{i}}\right]}}-1}. \qquad\qquad\qquad\qquad\qquad\qquad\qquad\qquad\qquad\qquad\qquad\;\; \label{eq:result_second_item_solving_mu_k}
\end{multline}
\end{figure*}
where $\boldsymbol{\Delta}\triangleq\mathbf{N}^{H}\mathbf{\boldsymbol{\Phi}}^{-1}\mathbf{M}$.
Likewise, substituting (\ref{eq:rewritten_V_tilde_k}) into $\mathrm{Tr}\left[\left(\mathbf{\widetilde{V}}_{k}^{\left(\mathit{n}+1\right)}\right)^{H}\mathbf{N}\right]$,
we have \eqref{eq:result_third_item_solving_mu_k}.
\begin{figure*}
\begin{multline}
\mathrm{Tr}\left[\left(\mathbf{\widetilde{V}}_{k}^{\left(\mathit{n}+1\right)}\right)^{H}\mathbf{N}\right]=\sum_{i=1}^{d_{k}}\boldsymbol{\Delta}_{\left[i,i\right]}^{H}+\sum_{i=1}^{d_{k}}\mathbf{\boldsymbol{\Lambda}_{\left[\mathrm{\mathit{i},\mathit{i}}\right]}}\frac{1-\left(\mathbf{\boldsymbol{\Gamma}}^{H}\boldsymbol{\Delta}^{H}\mathbf{\boldsymbol{\Gamma}}\right)_{\left[\mathrm{\mathit{i},\mathit{i}}\right]}}{\mathbf{\widetilde{\mu}}_{k}\mathbf{\boldsymbol{\Lambda}_{\left[\mathrm{\mathit{i},\mathit{i}}\right]}}-1}.\qquad\qquad\qquad\qquad\qquad\qquad\qquad\qquad\quad\label{eq:result_third_item_solving_mu_k}
\end{multline}
\end{figure*}

Finally, upon substituting (\ref{eq:result_first_item_solving_mu_k}),
(\ref{eq:result_second_item_solving_mu_k}) and (\ref{eq:result_third_item_solving_mu_k})
into the eavesdropper MSE constraint (\ref{eq:solving_mu_tilde_k_from_MSE_Eve}),
it is found that $\mathbf{\widetilde{\mu}}_{k}$ can be readily solved
from \eqref{eq:result_third_item_solving_mu_k-1},
\begin{figure*}
\begin{multline}
\mathbf{\mathrm{Tr}\left\{ \mathrm{\mathbf{E}_{Eve,Tx_{\mathit{k}}}\left[\mathbf{\widetilde{V}}_{k}^{\left(\mathit{n}+1\right)}\left(\widetilde{\mu}_{k}\right)\right]}\right\} }\\
=\sum_{i=1}^{d_{k}}\boldsymbol{\Omega}_{\left[i,i\right]}+\sum_{i=1}^{d_{k}}\boldsymbol{\Xi}_{[i,i]}-\sum_{i=1}^{d_{k}}\left(\mathbf{\boldsymbol{\Gamma}}^{H}\boldsymbol{\Xi}\mathbf{\boldsymbol{\Gamma}}\right)_{\left[i,i\right]}+\sum_{i=1}^{d_{k}}\boldsymbol{\Delta}_{\left[i,i\right]}+\sum_{i=1}^{d_{k}}\boldsymbol{\Delta}_{\left[i,i\right]}^{H}+\sum_{i=1}^{d_{k}}\frac{\mathbf{\boldsymbol{\Lambda}_{\left[\mathrm{\mathit{i},\mathit{i}}\right]}^{\mathrm{2}}}\left[\left(\mathbf{\boldsymbol{\Gamma}}^{H}\boldsymbol{\Delta}^{H}\mathbf{\boldsymbol{\Gamma}}\right)_{\left[i,i\right]}+\left(\mathbf{\boldsymbol{\Gamma}}^{H}\boldsymbol{\Delta}\mathbf{\boldsymbol{\Gamma}}\right)_{\left[i,i\right]}-1\right]\mathbf{\widetilde{\mu}}_{k}^{2}}{\left(\mathbf{\boldsymbol{\Lambda}_{\left[\mathrm{\mathit{i},\mathit{i}}\right]}}-\mathbf{\widetilde{\mu}}_{k}^{\left(n+1\right)}\right)^{2}}\qquad\qquad\qquad\quad\,\\
-\frac{\left\{ \mathbf{\boldsymbol{\Lambda}_{\left[\mathrm{\mathit{i},\mathit{i}}\right]}}\left[2-\left(\mathbf{\boldsymbol{\Gamma}}^{H}\boldsymbol{\Delta}^{H}\mathbf{\boldsymbol{\Gamma}}\right)_{\left[i,i\right]}-\left(\mathbf{\boldsymbol{\Gamma}}^{H}\boldsymbol{\Delta}\mathbf{\boldsymbol{\Gamma}}\right)_{\left[i,i\right]}\right]-\left(\mathbf{\boldsymbol{\Lambda}}\boldsymbol{\Pi}\right)_{\left[i,i\right]}-\left(\boldsymbol{\Pi}^{H}\mathbf{\boldsymbol{\Lambda}}\right)_{\left[i,i\right]}\right\} \mathbf{\widetilde{\mu}}_{k}}{\left(\mathbf{\boldsymbol{\Lambda}_{\left[\mathrm{\mathit{i},\mathit{i}}\right]}}-\mathbf{\widetilde{\mu}}_{k}^{\left(n+1\right)}\right)^{2}}-\frac{\left(\mathbf{\boldsymbol{\Gamma}}^{H}\boldsymbol{\Xi}\boldsymbol{\Gamma}\right)_{\left[i,i\right]}}{\left(\mathbf{\boldsymbol{\Lambda}_{\left[\mathrm{\mathit{i},\mathit{i}}\right]}}-\mathbf{\widetilde{\mu}}_{k}^{\left(n+1\right)}\right)^{2}}.\label{eq:result_third_item_solving_mu_k-1}
\end{multline}
\end{figure*}
where $\boldsymbol{\Xi}=\boldsymbol{\Delta}\boldsymbol{\Delta}^{H}$
and $\boldsymbol{\Pi}=\mathbf{\boldsymbol{\Gamma}}^{H}\boldsymbol{\Delta}^{H}\mathbf{\boldsymbol{\Gamma}}$.

\end{appendices}


\bibliography{Refs}

\noindent

\begin{IEEEbiography}[{\includegraphics[width=1in,height=1.25in,clip,keepaspectratio]{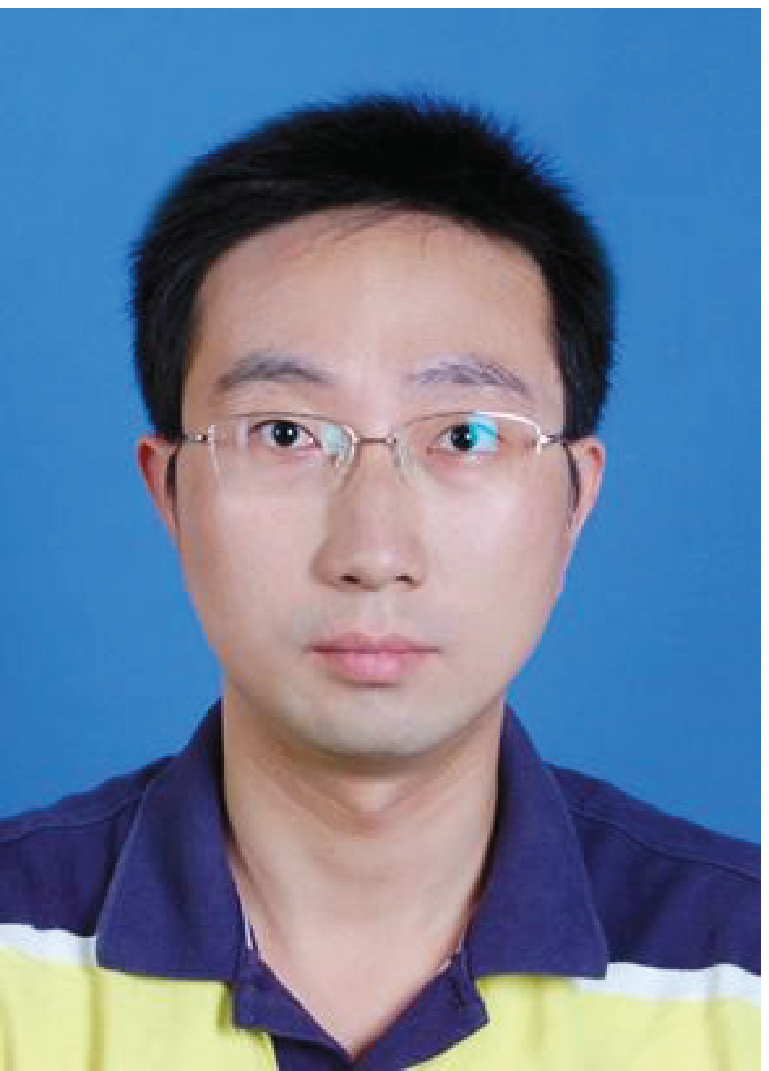}}] {Zhengmin Kong} received his B.Eng. degree and Ph.D. degree in 2003 and 2011, respectively, both from the School of Electronic Information and Communications,  Huazhong University of Science and Technology, Wuhan, China. From Sep. 2005 to Mar. 2011, he was with the Wuhan National Laboratory for Optoelectronics as a member of research staff and was involved in Beyond-3G (B3G) and UWB system design. He is currently a lecturer in the Department of Automation, Wuhan University. From Mar. 2014 to Mar. 2015, he was with the University of Southampton, UK as an academic visitor and investigated physical layer security and interference management techniques. His current research interests include wireless communications and signal processing, in particular physical layer security and interference management.
\end{IEEEbiography}

\begin{IEEEbiography}[{\includegraphics[width=1in,height=1.25in,clip,keepaspectratio]{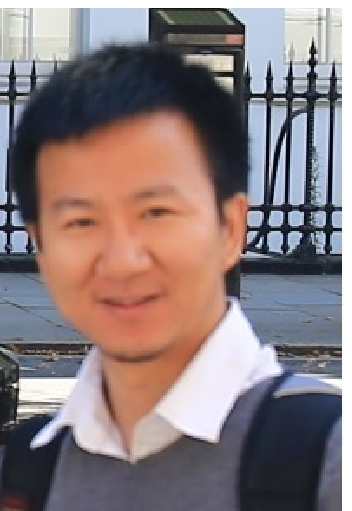}}] {Shaoshi Yang}
(S'09-M'13) received his B.Eng. degree in Information Engineering from Beijing University of Posts and Telecommunications (BUPT), Beijing, China in Jul. 2006, his first Ph.D. degree in Electronics and Electrical Engineering from University of Southampton, U.K. in Dec. 2013, and his second Ph.D. degree in Signal and Information Processing from BUPT in Mar. 2014. He is now working as a Postdoctoral Research Fellow in University of Southampton, U.K. From November 2008 to February 2009, he was an Intern Research Fellow with the Communications Technology Lab (CTL), Intel Labs, Beijing, China, where he focused on Channel Quality Indicator Channel (CQICH) design for mobile WiMAX (802.16m) standard. His research interests include MIMO signal processing, green radio, heterogeneous networks, cross-layer interference management, convex optimization and its applications. He has published in excess of 35 research papers on IEEE journals and conferences. 

Shaoshi has received a number of academic and research awards, including the prestigious Dean's Award for Early Career Research Excellence at University of Southampton, the PMC-Sierra Telecommunications Technology Paper Award at BUPT, the Electronics and Computer Science (ECS) Scholarship of University of Southampton, and the Best PhD Thesis Award of BUPT. He is a member of IEEE/IET, and a junior member of Isaac Newton Institute for Mathematical Sciences, Cambridge University, U.K. He also serves as a TPC member of several major IEEE conferences, including \textit{IEEE ICC, GLOBECOM, PIMRC, ICCVE, HPCC}, and as a Guest Associate Editor of \textit{IEEE Journal on Selected Areas in Communications.} (https://sites.google.com/site/shaoshiyang/) 
\end{IEEEbiography}

\begin{IEEEbiography}[{\includegraphics[width=1in,height=1.25in,clip,keepaspectratio]{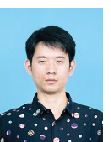}}] {Feilong Wu} received the B.S. degree in electronics information science and technology from Xi'an University of Science and Technology in July 2009. He is currently working toward the Ph.D. degree in communications and signal processing with the School of Electronics and Information Engineering, Xi'an Jiaotong University, Xi'an, China. From September 2013 to February 2015, he was a visiting Ph.D. student with the Southampton Wireless Research Group, University of Southampton, UK. He received the Best Paper Award of the International Conference on Wireless Communications and Signal Processing (WCSP) in 2011. His research interests include MIMO systems, wireless localization, and physical-layer security. 
\end{IEEEbiography}

\begin{IEEEbiography}[{\includegraphics[width=1in,height=1.25in,clip,keepaspectratio]{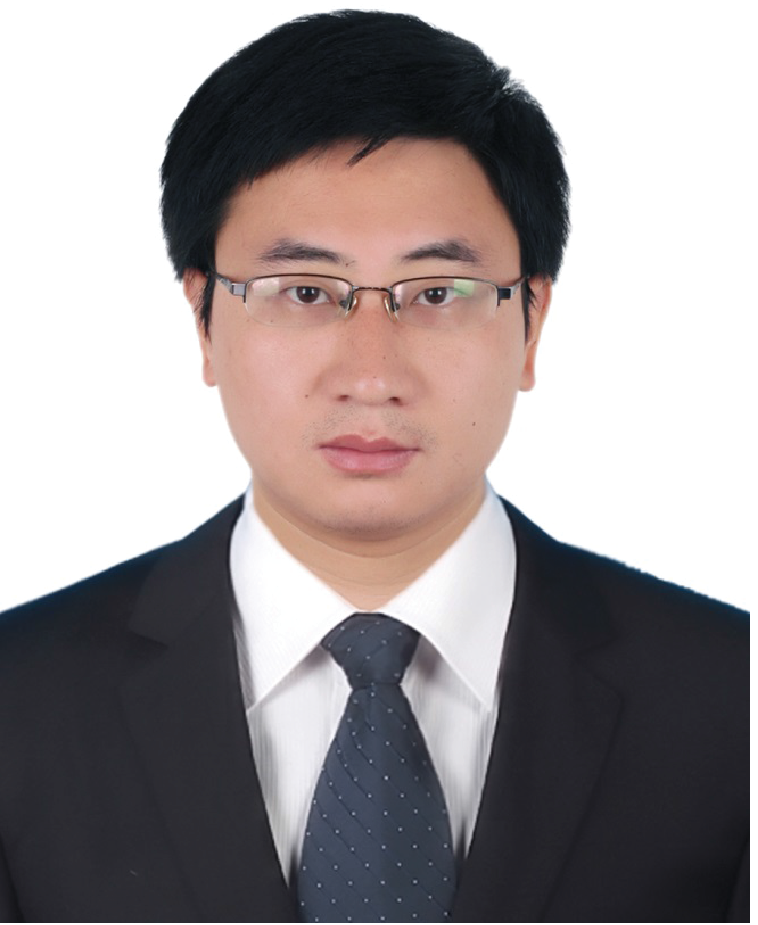}}] {Shixin Peng} received his Ph.D. degree in communications and information system from Huazhong University of Science and Technology, China, in 2015. He is currently a communication engineer at the State Grid Information and Communication Company of Hunan Electric Power Corp., Changsha, China. His research interests include interference management schemes in MIMO interference channels, capacity analysis in multiuser communication systems, 4G wireless communication systems, and signal processing.
\end{IEEEbiography}

\begin{IEEEbiography}[{\includegraphics[width=1in,height=1.25in,clip,keepaspectratio]{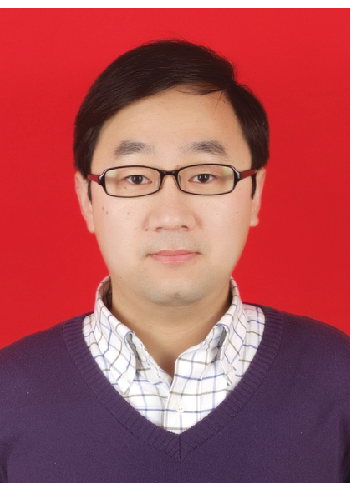}}] {Liang Zhong} received his Ph.D. degree in the Department of Electronics and Information Engineering, Huazhong University of Science and Technology, China, in 2013. He is currently a postdoctoral associate in the Department of Automation in Wuhan University, China. His research interests include interference management schemes in MIMO interference channels, wireless sensor networks, UWB communication systems, and signal processing.
\end{IEEEbiography}

\begin{IEEEbiography}[{\includegraphics[width=1in,height=1.25in,clip,keepaspectratio]{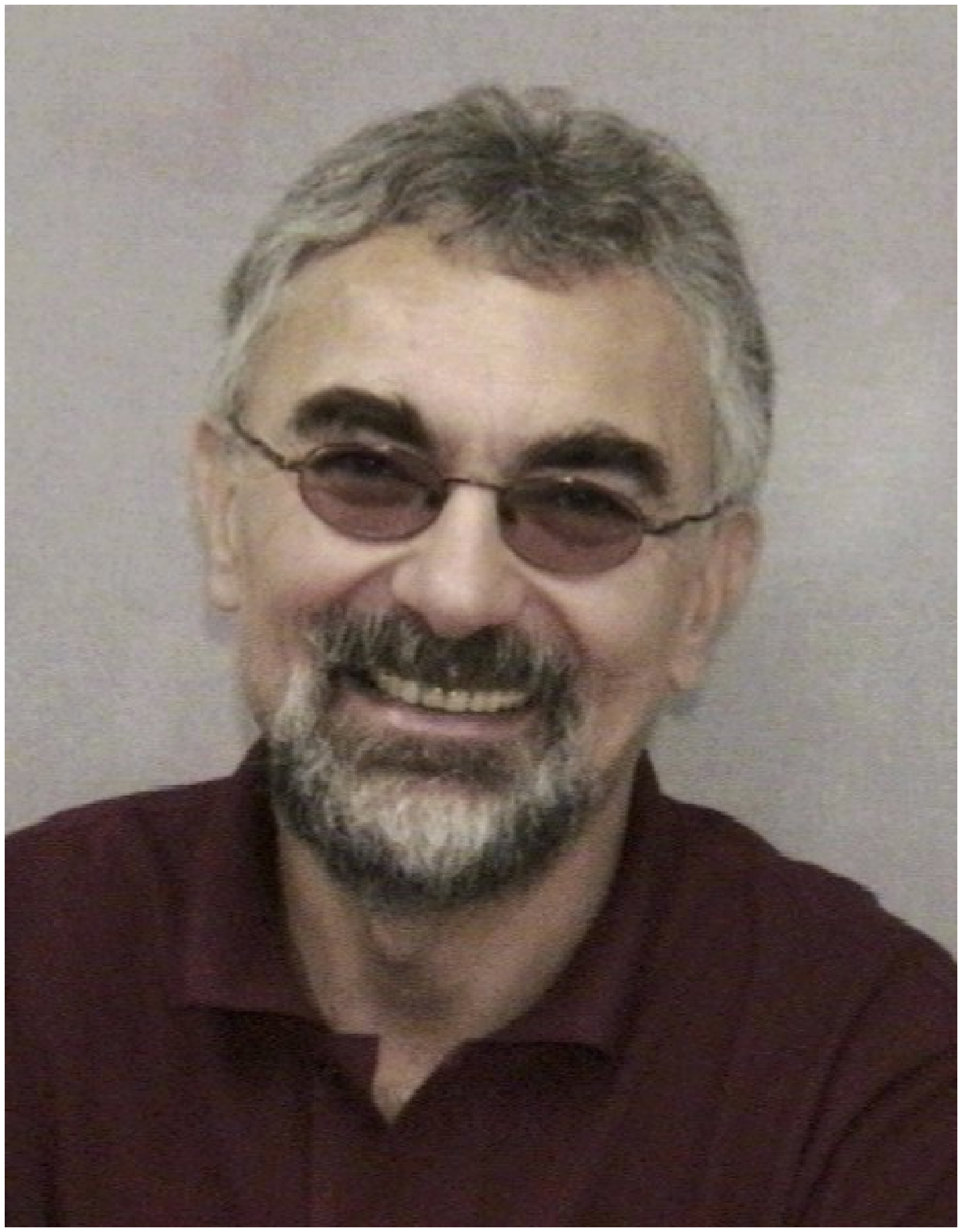}}] {Lajos Hanzo}
(M'91-SM'92-F'04) received his degree in electronics in
1976 and his doctorate in 1983.  In 2009 he was awarded the honorary
doctorate ``Doctor Honoris Causa'' by the Technical University of
Budapest.  During his 39-year career in telecommunications he has held
various research and academic posts in Hungary, Germany and the
UK. Since 1986 he has been with the School of Electronics and Computer
Science, University of Southampton, UK, where he holds the chair in
telecommunications.  He has successfully supervised 100+ PhD students,
co-authored 20 John Wiley/IEEE Press books on mobile radio
communications totalling in excess of 10 000 pages, published 1400+
research entries at IEEE Xplore, acted both as TPC and General Chair
of IEEE conferences, presented keynote lectures and has been awarded a
number of distinctions. Currently he is directing a 60-strong
academic research team, working on a range of research projects in the
field of wireless multimedia communications sponsored by industry, the
Engineering and Physical Sciences Research Council (EPSRC) UK, the
European Research Council's Advanced Fellow Grant and the Royal
Society's Wolfson Research Merit Award.  He is an enthusiastic
supporter of industrial and academic liaison and he offers a range of
industrial courses. 

Lajos is also a Fellow of the Royal Academy of Engineering, of the Institution
of Engineering and Technology (IET), and of the European Association for Signal
Processing (EURASIP). He is a Governor of the IEEE VTS.  During
2008 - 2012 he was the Editor-in-Chief of the IEEE Press and a Chaired
Professor also at Tsinghua University, Beijing. He 
has 22 000+ citations. For further information on research in progress and associated
publications please refer to http://www-mobile.ecs.soton.ac.uk 
\end{IEEEbiography}

\end{document}